*Spontaneous electric-polarization topology in confined ferroelectric nematics*

Jidan Yang[1]†, Yu Zou[1]†, Wentao Tang[1]†, Jinxing Li[1], Mingjun Huang[1,2]*, Satoshi Aya[1,2]*
† J.Y., Y.Z., W.T. contributed equally to this work.

[1] *South China Advanced Institute for Soft Matter Science and Technology (AISMST), School of Emergent Soft Matter, South China University of Technology, China.*
[2] *Guangdong Provincial Key Laboratory of Functional and Intelligent Hybrid Materials and Devices, South China University of Technology, Guangzhou 510640, China.*

*Abstract*

Topological spin and polar textures have fascinated people in different areas of physics and technologies. However, the observations are limited in magnetic and solid-state ferroelectric systems. Ferroelectric nematic is the first liquid-state ferroelectric that would carry many possibilities of spatially distributed polarization fields. Contrary to traditional magnetic or crystalline systems, anisotropic liquid crystal interactions can compete with the polarization counterparts, thereby setting a challenge in understating their interplays and the resultant topologies. Here, we discover chiral polarization meron-like structures during the emergence and growth of quasi-2D ferroelectric nematic domains, which are visualized by fluorescence confocal polarizing microscopy and second harmonic generation microscopies. Such micrometre-scale polarization textures are the modified electric variants of the magnetic merons. Unlike the conventional liquid crystal textures driven solely by the elasticity, the polarization field puts additional topological constraints, e.g., head-to-tail asymmetry, to the systems and results in a variety of previously unidentified polar topological patterns. The chirality can emerge spontaneously in polar textures and can be additionally biased by introducing chiral dopants. An extended mean-field modelling for the ferroelectric nematics reveals that the polarization strength of systems plays a dedicated role in determining polarization topology, providing a guide for exploring diverse polar textures in strongly-polarized liquid crystals.

*Introduction*

Topological defects arise when symmetry is broken. The complexity and diversity of the resulting spin or magnetic textures, such as skyrmions, merons, and the like, have attracted much attention for centuries[1-12]. In recent decades, it has been demonstrated that such vectorized fields with quasi-particle nature serve as an effective approach for integrating and storing magnetic, electronic, optical, and quantum information[13-19]. The findings trigger new challenges for the exploration of the unidentified topological fields, and their designability and controllability.

Quasi-particle topologies with vectorized fields in the context of (quasi-)2D systems, including rich species of vortices, skyrmions and merons, are usually characterized by three topological numbers: the Pontryagin (or skyrmion) number, vortex number and helical number. The Pontryagin number, $Q$, defines how many times a unit vector that defines the average orientation, the so-called director **n**, wraps the unit sphere [4, 20, 21], represented by

$$Q = \frac{1}{4\pi} \int \mathbf{n} \cdot (\partial_x \mathbf{n} \times \partial_y \mathbf{n}) \, dx dy. \tag{1}$$

$x$ and $y$ are the Cartesian coordinates. However, the usage of only $Q$ cannot fully describe various degenerated topological states. For example, the skyrmion and antiskyrmion cannot be distinguished only by $Q$. Thus, the additional vortex and helical numbers, $Q_v$ and $Q_h$, should be introduced to account for in-plane rotation and relative phase variation[4, 22]. By using these parameters, vectorized fields can be expressed

$$\mathbf{n}(\theta, \phi) = (\sin\theta \cos\phi, \sin\theta \sin\phi, \cos\theta)$$
$$= (\sin\theta \cos(Q_v\varphi + Q_h), \sin\theta \sin(Q_v\varphi + Q_h), \cos\theta). \tag{2}$$

where $\theta$ and $\phi$ are polar and azimuthal angles of director **n**, and $\varphi$ is the angle part of the polar coordinates for expressing the Cartesian coordinate $(x,y)$. For example, while an anticlockwise Bloch-type meron with an upward vector in the core exhibits a combination of the topological numbers $[Q, Q_v, Q_h] = [1/2, 1, \pi/2]$ (Fig. 1a), a divergent Neel-type anti-meron with a downward vector in the core shows $[Q, Q_v, Q_h] = [-1/2, 1, 0]$ (Fig. 1b). Though such topological textures have been widely found in magnetic systems, it is not until recently that several electric analogues were reported in multiferroic crystalline systems with superlattices[23-26]. This raises fundamental questions of whether more diverse polarization topological textures, beyond the magnetic counterparts, can emerge in broader physical systems, and how an additional interaction would bring unknown topologies into existence.

The recent discovery of a fluidic nematic-state ferroelectric, the so-called $N_F$ state[27-31], is the first instance of liquid-matter ferroelectric with high fluidity. Due to the liquid crystalline (LC) nature, their inherent elasticity and polarity, and multifaceted interactions with confined surfaces and responsivity to external fields, offer us fertile ground for exploring exotic states and emergent phenomena in soft matter physics. Therein, the polarization field is directly envisioned by second harmonic generation (SHG) microscopies thanks to its high nonlinear coefficient[32-39], and the system is easy to handle and manipulate. The advantages of the material category extend beyond the traditional magnetic systems and solid-state ferroelectrics if interplays between polarity and LC orientation happen. Here, we report on the discovery of various nontrivial polarization meron variants in the $N_F$ state. They exhibit spontaneous chirality symmetry breaking regardless of the achiral nature of the polar entities. By incorporating dipolar interaction, we extend the n-director-based mean-field theory with the Frank-Oseen potential[40-42] for describing the observed polarization topologies. The findings establish the fundamental physics of soft matter polarization textures and permit the exploration of the parameter space for their stabilization and diversification.

*Generation of quasi-2D polarization texture in $N_F$ droplets.* Traditional nematic (N) LCs that possess only the orientational order are apolar. The local average orientation of molecules is denoted by the director **n**. It exhibits the head-to-tail invariant nature in the apolar nematics, thereby leading to $\mathbf{n} \equiv -\mathbf{n}$ in the order parameter space of $\mathbf{S}^2/\mathbf{Z}_2$ in 3D (Fig. 1c)[43, 44]. This means the order parameter of the apolar nematic is a second-rank tensor. The $N_F$ state is the polar counterpart of the traditional nematics, which imposes an additional orientational requirement: $\mathbf{n} \neq -\mathbf{n}$ in the order parameter space of $\mathbf{S}^2$ where the head and tail of the director are differentiable (Fig. 1d)[4, 20, 45, 46]. This polar nature of the orientational directionality is expected to give much more diverse topologies. As seen in the following, even limiting to trivial situations, the emergence of the vectorized polarization at least doubles the variety of the topology compared with the apolar N state. As some explicit examples, we consider correspondence between some topological structures in the apolar and polar systems for both 2D and 3D cases (Figs. 1e-p). In 2D case, the apolar and polar counterparts stay in the order parameter spaces of $\mathbf{S}^1/\mathbf{Z}_2$ and $\mathbf{S}^1$, respectively. The topological defects in the two types of director fields are points in the cores for both the apolar and polar systems (Figs. 1e-j). In 3D case under a spherical confinement, the order parameter spaces become $\mathbf{S}^2/\mathbf{Z}_2$ and $\mathbf{S}^2$. Extending the 2D radial director fields (Fig. 1e) filling the sphere, the +1 radial defects can be trivially transformed to hedgehog-type defects under the homeotropic boundary condition (Fig. 1k). In the polar versions, the hedgehog branches into two distinct states that correspond to the radial divergent and convergent polarization fields (Figs. 1l,m). On the other hand, the 3D counterparts of the 2D concentric +1 defects under the planar boundary condition remain diverse. The direct space-filling of the concentric +1 defects in 3D sphere can possibly carry line disclination (Figs. 1n-p) or be nontrivial. The topological complexity will be affected by the competition of elasticity and polar interactions as discussed below.

In this study, we synthesized a new $N_F$ material with a large dipole moment of ~13.6 D, named RM-OC$_2$ (Fig. 1d and Fig. S1). The structure is based on RM734[29, 47] with additions of a nitro and an ethoxy group at the third and second benzene rings, respectively. Such a molecular design leads to a larger lateral bulkiness and a more oblique molecular dipole with respect to the molecular long axis. Since the molecular shape significantly deviates from the ideal rod, we anticipated the traditional nematic state is not preferred. Instead, anisotropic polar interactions serve as the main driving force for generating an anisotropic orientation and thus stabilizing a polarity-driven nematic phase, i.e., the $N_F$ state. Fitting to our expectation, though most of the previously-reported $N_F$ states emerge from the high-temperature apolar N state[27, 29, 48, 49], RM-OC$_2$ exhibits a direct transition from the isotropic liquid (Iso) to the $N_F$ state upon cooling with the phase sequence of Iso-[64.6 °C]-$N_F$ (Figs. 2a-d). On the heating, the material shows several recrystallization processes i.e. $N_F$-[67.1 °C]-Iso-[82.1 °C]-unknown crystal X1-[105~120 °C]-unknown crystal X2-[140 °C]-Iso on heating (Figs. S1b-j). The recrystallization processes suggest the Iso phase is metastable with respect to some crystalline forms in the range of 80~150 °C. Like the other emerging $N_F$ materials[48-50], both the dielectric constant and second harmonic generation (SHG) signal dramatically increase upon transitioning into the $N_F$ state (Fig. 2a). The direct Iso-$N_F$ transition pathway modifies the $N_F$ topology because of the evolution of the polarization field from disorder states. In the previously reported cases with the N-$N_F$ transition[27, 29, 48-50], the low-temperature $N_F$ polarization (or director) field develops by breaking the head-to-tail invariance of **n** of the high-temperature apolar N state. It accompanies a collective flipping process of the molecular polarity to a preferred direction over a length scale larger than micron-scale[27, 47, 51]. Therefore, the orientational direction of the molecular polarity is affected by the orientational pattern of the high temperature apolar N state. Upon the direct Iso-$N_F$ transition, the polar 'spherulite' domains appear from the Iso background through nucleation and growth processes (Figs. 2b,c), finalizing into the band-like texture with disclination lines (Fig. 2d and Figs. S14-15). Such a phase evolutional pathway allows us to track the pure emerging polarization field from a disordered state.

*Observation of polarization meron variants*. In $N_F$ droplets, the texture (Figs. 3a-c and Fig. S2) cannot be brought into an extinction status under the crossed polarizers, especially in the center. These observations suggest the $N_F$ droplets carry a spontaneous twist along the depth direction, which causes the optical rotation effects. This was immediately confirmed in detail by the polarized light microscopy (PLM) decrossing technique. The clockwise (Fig. 3a) or counter-clockwise (Fig. 3c) rotation of the analyzer distinguishes two different types of textures: at specific decrossing conditions, one type turns to a dark state (i.e. extinction) and the other remains bright. This indicates the two types have opposite handedness. With the insertion of a quarter-wave plate, it seems that the two types of patterns correspond to concentric director fields with left hand (Figs. 3d,e) and right hand (Figs. 3f,g) spirals. To obtain 3D orientational information, we employ the fluorescence confocal polarizing microscopy (FCPM) and deduce the director field[40, 52] (Methods). In principle, the polarizing fluorescence process gives rise to a $\cos^2 \beta$ orientational dependence of the fluorescence signal, where $\beta$ is the angle between the incident linear polarization and local director. Also, the out-of-plane of director tilting causes the decrease of fluorescence. Thus, analysing the spatial distribution of fluorescence intensity at proper combination of polarization conditions allows us to calculate the director field (Methods). Technically, while the linear polarization pumping makes the in-plane and out-of-plane director tilting undifferentiable, the addition of experiment with a circular polarized pumping light at the same field-of-view extracts the pure information of the out-of-plane tilting[52]. Therefore, the in-plane and out-of-plane director tilting information are uncoupled and accessible. Worth noting, it is exclusively possible to know the average molecular orientation, but the polarization information is still missing. Figures 3h-k demonstrate the FCPM images of the two types of patterns. It confirms that the $N_F$ droplets confined in the thin LC cells show cylindrical shape with some curvatures. From the 2D cross sections at various depths (Fig. S3), it is seen that, consistent with the PLM observation, two types of director fields spontaneously twist like spirals. The 3D orientational information for deriving the overall topology can be further deduced by calculating the 3D distribution of director orientation. As seen from the FCPM under a circular polarized pumping light, a decrease of fluorescence occurs near the droplet center, while the other areas show a constant fluorescence intensity (Fig. S4c). In the depth cross section profile, the fluorescence profile in the droplet center appears as a broaden black cylinder along the surface normal. This suggests the director field exhibits either a significant out-of-plane tilting towards the center (i.e. homeotropic-like in the center) or a disclination line. The possibility of the existence of a line disclination is excluded by experiment. The fluorescence profile of a designed line disclination by photo-pattern is thin and sharp (Fig. S4a), which disagrees with the broaden black cylinder in the fluorescence profile of droplets (Fig. S4b). This suggests that the broad range of light extinction corresponds to the out-of-plane tilting of director. Therefore, the observed topology is limited to either a meron-like structure or an escaped vortex structure as further discussed later in Discussion (Fig. 6 and Figs. S9-10). Additional simulations of the fluorescence profile of these two possible structures deduced from numerical structural relaxation (Methods, Discussion and Figs. S5a-c) further confirm the topology of the droplet. It is clear that the signal profile of meron-like structures (Fig. S5b) is consistent with our experimental result (Fig. S5d), suggesting the observed topology is meron-like structures (Figs. 3l,m) if such a polarization field exists in addition to the orientational field.

Unsimilar to the traditional head-to-tail-equivalent director field, the polarization field should be additionally addressed by SHG microscopy methods to investigate the head-to-tail-inequivalent polarization field[36, 48]. Figure 4a demonstrates a large-area SHG confocal polarizing microscopy (SHG-CPM) image of the emerging $N_F$ droplets in the cell midplane[53]. The large contrast between the $N_F$ droplets and apolar background signals visualizes the polarization field. Distinct to the polarizing fluorescence process with a $\cos^2 \beta$ orientational dependence, the SHG process exhibits a $\cos^4 \beta$ orientational dependence of the SH signal. The SH signal is the largest/smallest when the polarization of excitation light is (anti)parallel/perpendicular to the local polarization. It is clearly seen that the coexistence of the two types of director fields as deduced from FCPM emits a strong SH signal. Worth noting, the signal distributions in FCPM and SHG images are well consistent, suggesting the local polarization is parallel to the director (Figs. 3h,j and 4b-g). In the normal SHG measurement, while the existence of the polar symmetry is probed, the head and tail of the polarization cannot be differentiated. It means that the information on the vectorized orientation of the polarization is overlooked. SHG interferometry (SHG-I) microscopy is an extended SHG microscopy technique that 'sees' polarization vector by differentiating the relative difference of the optical phase information[33, 36, 54] (Methods). Figures 4h-k confirm four types of distinct polarized SHG-I images. The four types of polarization meron-like structures could not be resolved by either FCPM or SHG-CPM where only two types of patterns have been seen (Figs. 3h-k and 4a). They correspond to four combinations of left or right hand meron-like structures with divergent or convergent polarization fields (Figs. 4p-s). The chiral symmetry breaking under the confinement occurs due to the delicate balance between the elastic and polar interactions as explained in Discussion. They show almost equal number of appearances (inset in Fig. 4a), indicating that the handedness of the spontaneous chirality in the polarization meron-like structures is induced by accident.

*Chirality biasing of meron invariants.* Introducing chiral agents is a general way for producing an additional twist and biasing the handedness. Here, for retaining the same topological structures, we add a very small amount of two commercial chiral dopants, right-handed R811 or left-handed S811, into the system to see the degree of the symmetry breaking. As shown in Figure S6, above a threshold weight percentage, > 0.1 wt%, the handedness of the system is totally biased to that of the chiral dopants. Combining PLM, SHG-CPM and CFPM, the system exhibits only two types of the polarization meron-like structures (Fig. 5 and Fig. S7), surviving from the original four types of polarization meron-like structures (Figs. 4p-s). The polarization fields are again clarified by SHG-I images: convergent or divergent polarization field for a certain handedness of twist determined by the chiral dopant (Fig. S8). These observations verify that the system stabilized one of the twisted handedness under the action of the chiral dopants.

*Discussion*

*Effect of polar interactions on polarization topology.* We discuss what leads to the spontaneous chirality in the achiral $N_F$ system. Due to the emergent polarization field, polar interactions as well as the traditional elastic contributions are vital for dedicating the global topology. Therefore, we consider an extended Frank-Oseen free energy function of the system by incorporating dipolar interactions besides the Landau, elastic, surface anchoring and defect core energies (Methods). We uncover five possible polarization states in the curved cylindrically confined space: polarization vortex (Fig. 6a), escaped vortex (Fig. 6b), concentric meron-like structure (C-meron; Fig. 6c), divergent meron-like structures (D-meron, including the convergent, so negatively divergent, meron-like structure; Fig. 6d) and bipolar structure (Fig. 6e). In the simplest polarization vortex (Fig. 6a), the polarization field arranges in a pure concentric manner and lies on the sample plane (i.e. wrapping the equator of the order parameter space $S^2$ sphere). This leads to a disclination line in the center. In the escaped vortex (Fig. 6b), the polarization continuously flips along the depth direction, enabling the disclination line of simple polarization vortex to escape onto the middle plane. Now, the polarizations span over the whole $S^2$ sphere. Since the disclination line of the simple polarization vortex costs much larger energy than that of the point defect of the escaped vortex in our experimental geometry (see Methods for free energy descriptions), the escaped vortex is more stable than the simple polarization vortex in most of the simulation conditions (Fig. 6g and Figs. S11-13). Meron-like structures are distinct from the vortices because they do not exhibit singularities. C-meron demonstrates a pure concentric polarization field except for the core region (Figs. 6c). D-meron exhibits additional directional splay-bend distortions (Figs. 6d), which is consistent with the experimental observation of the in-plane handedness of meron-like structures. The bipolar structure possesses two boojum defects on the north and south poles (Figs. 6e), which has been well studied in the classical apolar nematic droplets[55-58]. Details of the two-dimensional distribution of polarization fields at different sample depth for all the structures is shown in Fig. 6f.

It is important to mention that the experimentally-observed D-meron like structure appears only when $K_{11}$ satisfies proper conditions: 0.2 pN < $K_{11}$ < 0.8 pN. The variation of $K_{22}$ or $K_{33}$ with respect to other elastic moduli seems to have less potential to induce D-meron (Figs. S11-13). Herein, we demonstrate a representative simulated state diagram by varying the effective polarization strength $P_0$ and the elastic anisotropy (Fig. 6g). The elastic anisotropy is defined as a ratio of the splay to bend elasticity $K_{11}/K_{33}$ (assuming the twist and bend elastic modulus are 0.8 pN and 2.0 pN, respectively, as are close to the reported values in the vicinity of the N-N$_F$ phase transition[47]). When the polarity is absent or small in the system (so $P_0 < 2.5 \times 10^{-4}$ C m$^{-2}$), the bipolar structure with a large portion of splay deformation is dominant in small $K_{11}/K_{33}$ values ($K_{11}/K_{33} < 0.4$). With increasing $K_{11}/K_{33}$ (e.g., path I in Fig. 6g), so increasing the penalty of the splay deformation[42, 55, 56, 59], the bipolar structure becomes unstable and is replaced by the escaped vortex-like apolar director field. Further increasing the $K_{11}/K_{33}$ will cause a further transition to an apolar variant of the C-meron. We note that, keeping mind that our system has a curved cylindrical shape like a cylinder LC cell, the corresponding apolar textures in the sphere droplet geometry with tangential anchoring have been well studied in traditional nematic materials by simulations and experiments[55-57, 59-61]. The bipolar structure is easier to form in spherical confined space. It is reported that when $K_{11}/K_{33} \leq 1$, the bipolar structure with two boojums defects can be observed in spherical nematic droplets[60]. Especially, when $K_{11} \geq K_{22} + 0.43 K_{33}$, this typical bipolar structure will distort and change to a twisted bipolar structure[56]. If $K_{11}$ is further increased, the most stable energy structure in the spherical droplet is no longer the apolar bipolar structure, but the apolar concentric structure. This tendency in the reports is consistent with the result for $P_0 = 0$ in our cylindrical droplets.

Introducing polarity into the system dramatically changes the director field. In small $K_{11}/K_{33}$ range, when the polarization strength $P_0$ increases (e.g., path II in Fig. 6g), due to the preference of antiparallel packing of local polarizations by the enhanced dipolar interaction, the bipolar or escaped vortex director field changes to the defect-less D-meron. In larger $K_{11}/K_{33}$ range ($K_{11}/K_{33} > 0.4$), the polar version of C-meron with a pure bend (so concentric) replaces D-meron. For C-meron, because the topology is invariant in the depth direction, the topology only carries two possibilities of $[Q, Q_v, Q_h] = [1/2, +1, \pm\pi/2]$, i.e. clockwise or anticlockwise. For both cases, due to the absence of in-plane and out-of-plane twisting structures, they cannot show the non-extinct textures as observed in the experiment (Fig. 3). On the other hand, for D-meron, an in-plane splay-bend deformation and an accompanying spiral streamline of polarization from the center exist. This results in eight species of topology with distinct combinations of topological numbers, $[Q, Q_v, Q_h] = [\pm 1/2, +1, \pm\pi/2]$ and clockwise/counter-clockwise spirals. However, the polarization fields of a meron and an anti-meron with opposite handedness are degenerate in experiments since the current nonlinear optical microscopy have no axial resolution for differentiating the polarization states along light propagation direction, i.e. up or down along the viewing direction (Fig. S9-10). Thus, only four types of meron-like structures are experimentally observed (Fig. 4l-o). The appearance necessitates the system possessing a strong polarization strength over the transition threshold, i.e. $P_0 > 2.5 \times 10^{-4}$ C m$^{-2}$.

Summarizing, the real-space observation of polarization meron variants in the fluidic LC state resembles that of magnetic and spin systems. The results suggest the importance of competition between polar and elastic anisotropies. Our findings shed new light on understanding the complex relationship between polar interactions and diverse liquid crystalline orderings, and open the door for the design and engineering of polarization topologies in liquid matter systems.

*Methods*

*Sample Preparation.* RM-OC$_2$ (Fig. S1) is synthesized in the laboratory (Materials and Methods in Supplementary Information). The helielectric materials are prepared by mixing a chiral dopant, (R)-2-Octyl 4-[4-(Hexyloxy)benzoyloxy]benzoate (R811, TCI) or (S)-2-Octyl 4-[4-(Hexyloxy)benzoyloxy]benzoate (S811, TCI) with RM-OC$_2$. The concentrations of the chiral dopants are adjusted in the range of 0.05-1 wt%.

*Sample Preparation in LC slab.* The samples are introduced to homemade or commercial LC slabs with controlled thickness and planar alignment conditions. The glasses are treated with KPI-3000 (Shenzhen Haihao Technology Co. Ltd.). Homemade planar slabs with thickness in the range of 3.2-10 μm are used for the observation of PLM texture. Homemade 5-μm thick slabs with planar alignment are used for measuring the SH and SHG interference (SHG-I) signals. Homemade 5-μm thick slabs with planar alignment are used for the FCPM observations. We use homemade 10-μm thick cells with Cr and Au coated layers to conduct the dielectric properties characterization. No further surface treatment is made on the metal layers.

*Fluorescence confocal polarizing microscopy.* FCPM observation is performed using an inverted microscope equipped with a halogen illuminator (Zeiss HAL 100 illuminator). The temperature of samples is controlled by a homemade temperature sink and controller. A 63x oil-immersion objective lens with numerical aperture NA=1.4 and a 40x objective lens with numerical aperture NA=0.95 are used. We use fluorescent dye 9-(Diethylamino)-5H-benzo[a]phenoxazin-5-one (Nile Red, Sigma-Aldrich) which is doped into the RM-OC$_2$ or RM-OC$_2$R811(S811) mixture. The concentration of the fluorescent dye is 0.05 wt%. The Nile Red molecules align parallel to the LC director. We use a linear polarization for the excitation of the dye. The wavelength of the incident laser is 514 nm. The detected fluorescence intensity depends on the angle α between the polarization direction and the director. Thus, by collecting the fluorescence intensity at four polarization conditions of $α_1$=0, $α_2$=π/4, $α_3$=π/2 and $α_4$=3π/4, we determine the orientation of local directors $\mathbf{n} = (\cos\psi\cos\theta, \sin\psi\cos\theta, \sin\theta)$. $\psi$ is the azimuth angle measured from the $x$ axis in the xy-plane and $\theta$ the out-of-plane polar angle measured from the xy-plane. The detected fluorescence intensity follows the equation:

$$I_k = I_{\text{back}} + I_{\text{norm}}\cos^4(\psi - k\pi/4)\cos^4\theta. \tag{3}$$

$k$ takes a range from 0 to 3 and the angle between polarization directions and $x$ axis is $k\pi/4$. $I_{\text{back}}$

is the background intensity and $I_{\text{norm}}$ the normalized fluorescence. $I_k$ is the transmittance at the polarization condition of $\alpha_k$. The angle $\psi$ is thereby calculated by $\psi = \frac{1}{2}\tan^{-1}\frac{I_1-I_3}{I_0-I_2}$. Then, $\theta$ is determined by substituting the calculated $\psi$ into Eq. (3). However, after substitution, the sign of $\theta$ remains unknown. Assuming the polarization field is continuous as also confirmed by SHG-I microscopy, we can deduce the angle $\theta$ in 3D. The angle $\psi$ lies in the range of $[0,\pi]$, which means FCPM does not differentiate the head or tail of polarization. As a result, FCPM gives only the non-vectorized orientational field. The polarization field is deduced by combining SHG-I data and the 3D non-vectorized orientational field probed by FCPM as explained below.

*SHG and SHG interference (SHG-I) measurement.* We use a fundamental beam from a Q-switched pulsed laser (MPL-III-1064-20μJ) with a central wavelength of 1064 nm, maximum power of 200 mW, pulse duration of 5 ns, and 100 Hz repetition. The fundamental beam is set to be polarized and directed into LC cells. The schematics of the optical system can be found in ref. 42. The SH light is detected in the transmission geometry by a photomultiplier tube (DH-PMT-D100V, Daheng Optics) or a scientific CMOS camera (Zyla-4.2P-USB3, Andor). For the temperature scanning, the SH signal is recorded every 1 °C under the control of a home-made Labview program. For the SHG interferometry, we use two fused silica plates for generating the phase difference between the SH signal from the sample and the reference Y-cut quartz plate. A fused silica plates are rotated in opposite directions to avoid the variation of the optical path during their rotation. With larger rotating angle of the fused silica plates, the induced phase difference is larger. For the SHG-I microscopy, we choose several interference conditions for the best imaging contrast by finding the corresponding rotating angles of the fused silica plates. We also use a commercial inverted microscope (LSM880, Zeiss) equipped with a femtosecond laser (Integrated pre-compensation 80MHz, Vision II, Chameleon system) for high-speed large area visualization. The wavelength of the incident laser is 800 nm.

*Dielectric spectroscopy.* The dielectric spectroscopy is conducted by using an LCR meter (4284A, Agilent). The data collections of the frequency and temperature sweeping are automated by a home-made software written in Labview.

*Numerical Modelling.* As described in the main text, we consider the free energy density contributions from the dipolar interaction ($f_{\text{d}}$) besides the Landau ($f_{\text{L}}$), elastic ($f_{\text{e}}$), surface anchoring ($f_{\text{s}}$) and defect core ($F_{\text{core}}$) energies:

$$F = \int (f_{\text{L}} + f_{\text{e}} + f_{\text{d}})\,\mathrm{d}V + \int_{\Omega} f_{\text{s}}\,\mathrm{d}\Omega + F_{\text{core}}, \tag{4}$$

$$f_{\text{L}} = \frac{a}{2}|\mathbf{P}|^2 + \frac{b}{2}|\mathbf{P}|^4, \tag{5}$$

$$f_{\text{e}} = \tfrac{1}{2}K_{11}(div\,\mathbf{n})^2 + \tfrac{1}{2}K_{22}\bigl(\mathbf{n}\cdot(curl\,\mathbf{n})\bigr)^2 + \tfrac{1}{2}K_{33}\bigl(\mathbf{n}\times(curl\,\mathbf{n})\bigr)^2, \tag{6}$$

$$f_\mathrm{d} = \frac{1}{8\pi\varepsilon_0\varepsilon_r} \int \left\{ \frac{\mathbf{P}(\mathbf{r}')\cdot\mathbf{P}(\mathbf{r})}{|\mathbf{r}-\mathbf{r}'|^3} - \frac{3[\mathbf{P}(\mathbf{r}')\cdot(\mathbf{r}-\mathbf{r}')][\mathbf{P}(\mathbf{r})\cdot(\mathbf{r}-\mathbf{r}')]}{|\mathbf{r}-\mathbf{r}'|^5} \right\} \mathrm{d}V, \quad (7)$$

$$f_\mathrm{s} = \frac{1}{2} W_\mathrm{S} (\cos\theta - \mathbf{n}\cdot\mathbf{v})^2. \quad (8)$$

The Landau energy, $f_\mathrm{L}$, is summed up to the fourth order, describing the stability of the $N_F$ state. Namely, the deeper potential-well at a finite polarization $\mathbf{P} = P_0\mathbf{n}$ corresponds to a more stable ferroelectric state. $P_0$ is the polarization strength of polar molecules. The elastic energy term, $f_\mathrm{e}$, treats the traditional nematic elastic functionals, where only the terms that are quadratic in $\mathbf{n}$ are used[49, 62]. It penalizes the splay, twist and bend elastic deformations through the elastic constants of $K_{11}$, $K_{22}$ and $K_{33}$. The dipolar interaction, $f_\mathrm{d}$, describes the electrostatic contributions from all the interacting polar bodies in the system, which includes both the space charge arising from the splay polarization field, $\nabla \cdot \mathbf{P}(\mathbf{r})$[63], and boundary charge due to the discontinuity of the polarization on the interface. $\varepsilon_0$ is the vacuum permittivity, $8.85 \times 10^{-12}$ C V$^{-1}$ m$^{-1}$. $\varepsilon_r$ is the relative permittivity, which is set to be $10^4$ in our simulations. The value is comparable to the value reported in the literatures[27, 48]. Here, due to a constant of $\varepsilon_r$ is used for calculation, though the contribution of the variation of the polarization field is considered, the orientation-dependent dielectric interaction is not fully accounted in the present form. $W_S$, $\mathbf{v}$ and $\theta$ in the surface anchoring energy are the anchoring coefficient, the normal vector of the droplet surface, and the deviation angle between the polarization and surface normal vectors. Due to the surface charge and surface tension effects, we assume that the Iso-$N_F$ and $N_F$-glass interfaces pose a degenerate anchoring, where director field lies in the interface plane but without specific azimuthal orientational constraint. To reasonably consider this, we employ the degenerate planar anchoring condition[64], i.e. $\theta = \pi/2$ and $W_\mathrm{s} = 10^{-6}$ J/m$^2$. According to eq. (4), the free energy terms are integrated using the finite volume and surface area elements $\mathrm{d}V = \mathrm{d}x\mathrm{d}y\mathrm{d}z$ and $\mathrm{d}\Omega = \mathrm{d}x\mathrm{d}y$, respectively. The defect core free energy, $F_\mathrm{core}$, for three defect-holding polarization states, i.e. simple polarization vortex, escaped polarization vortex and bipolar structure, are calculated as:

$F_\mathrm{core} = \pi K L \ln(R_1/a)$ (for simple polarization vortex)[42],

$F_\mathrm{core} = 8/3 \pi K R_2$ (for escaped polarization vortex)[65],

$F_\mathrm{core} = 8\pi K R_2$ (for bipolar structure)[65].

$K$ is taken as the average of the splay, twist and bend elastic moduli. $R_1$ and $R_2$ represent the radius of line defects and point defects, respectively. $L$ is the length of line defects. $a$ is the average molecular size. The polar vortex contains a line defect (Fig. 6a), whose length is equal to the sample thickness. Under this condition, the calculated energy is comparable to the energy of the line defect reported in apolar LC systems[42]. The escaped vortex and bipolar structures carry point defects (Fig. 6b and Fig. 6e). The core radius of the defects is assumed to be comparable to the nematic correlation length, i.e. $R_1 = R_2 = 15$ nm[66]. In the simulation, we create the 3D curved cylindrical geometries in Creo Parametric 4.0, which is nearly consistent with the FCPM experimental observation (Figure S3). The 3D models are built to have orthogonal meshes and are used to run the n-director simulations. We set the maximum and minimum diameters of the curved cylinder to be 10 and 5 μm, respectively, which locates in the middle plane of droplets and at the interface between the glass and LCs. The sample thickness is set to be 5 μm. We perform the relaxation minimization of the corresponding Ginzburg–Landau functionals based on the finite element difference method (within a 3D space of 14 μm × 14 μm × 17 μm). The 3D space is orthogonally divided by grids with the grid spacing in the range from 10 nm to 500 nm. In this range, we make sure that the numerically-calculated topology is consistent, where the topological details do not depend on the size of the grid spacing. To investigate how the strength of polarity and elasticities would affect the topological nature, we vary $P_0$ and the ratio of elasticities, $K_{11}/K_{33}$, for the simulations. Note that, unlike the simulation by using the Landau–de Gennes Q-tensor[42] where the order parameter can change during orientational relaxation, the predesignated $P_0$ for each simulation condition in the present free-energy context is unchanged during simulation runs since both the polarity strength (so the order parameter of polarization) and the nematic order parameter are constants, i.e. $|\mathbf{P}| = P_0$ and $|\mathbf{n}| = 1$. Under this circumstance, the Landau energy in eq. (5) is a constant for a chosen combination of ($P_0$, $K_{11}/K_{33}$). This means that the term will not affect what topological states would be chosen. Therefore, in our simulation, the Landau term is discarded. Considering, in our experiment, the $N_F$ droplets appear from the disordered Iso state, we set the initial condition of the director field to be random for most cases. For the conditions at $P_0 < 1 \times 10^{-4}$ C m$^{-2}$ and $K_{11}/K_{33} > 0.5$, the simulations end to ideal relaxed topologies as shown in Figure 6g. However, when the polarization strength is large ($P_0 > \sim 1 \times 10^{-4}$ C m$^{-2}$) and $K_{11}$ is small ($K_{11}/K_{33} < \sim 0.5$), which corresponds to the left corner of the state diagram, a random initial condition would result in somehow disordered results by some trapped defects. For these situations, we use the relaxed structures of the simulation at slightly smaller $P_0$ or larger $K_{11}$ as initial condition for the numerical relaxations. We relax the systems to the equilibrium state, where the free energy is the global minimum. We note that the initial condition does not change the equilibrium topology.

*Reference*

1. Rossler UK, Bogdanov AN, Pfleiderer C. Spontaneous skyrmion ground states in magnetic metals. *Nature* **442**, 797-801 (2006).
2. Soumyanarayanan A, Reyren N, Fert A, Panagopoulos C. Emergent phenomena induced by spin-orbit coupling at surfaces and interfaces. *Nature* **539**, 509-517 (2016).
3. Mühlbauer S*, et al.* Skyrmion Lattice in a Chiral Magnet. *Science* **323**, 915-919 (2009).

4.  Nagaosa N, Tokura Y. Topological properties and dynamics of magnetic skyrmions. *Nat Nanotechnol* **8**, 899-911 (2013).
5.  Fukuda J, Zumer S. Quasi-two-dimensional Skyrmion lattices in a chiral nematic liquid crystal. *Nat Commun* **2**, 246 (2011).
6.  Jiang W, Chen G, Liu K, Zang J, te Velthuis SGE, Hoffmann A. Skyrmions in magnetic multilayers. *Phys Rep* **704**, 1-49 (2017).
7.  Yu XZ, *et al.* Transformation between meron and skyrmion topological spin textures in a chiral magnet. *Nature* **564**, 95-98 (2018).
8.  Augustin M, Jenkins S, Evans RFL, Novoselov KS, Santos EJG. Properties and dynamics of meron topological spin textures in the two-dimensional magnet $CrCl_3$. *Nat Commun* **12**, 185 (2021).
9.  Thiaville A, Miltat J. Topology and Magnetic Domain Walls. In: *Topology in Magnetism* (eds Zang J, Cros V, Hoffmann A). Springer International Publishing (2018).
10. Wang XS, Wang XR. Topology in Magnetism. In: *Chirality, Magnetism and Magnetoelectricity: Separate Phenomena and Joint Effects in Metamaterial Structures* (ed Kamenetskii E). Springer International Publishing (2021).
11. Tokura Y, Kanazawa N. Magnetic Skyrmion Materials. *Chem Rev* **121**, 2857-2897 (2021).
12. Smalyukh I. Review: Knots and other new topological effects in liquid crystals and colloids. *Rep Prog Phys* **83**, 106601 (2020).
13. Tomasello R, Martinez E, Zivieri R, Torres L, Carpentieri M, Finocchio G. A strategy for the design of skyrmion racetrack memories. *Sci Rep* **4**, 6784 (2014).
14. Parkin SSP, Hayashi M, Thomas L. Magnetic Domain-Wall Racetrack Memory. *Science* **320**, 190-194 (2008).
15. Fert A, Cros V, Sampaio J. Skyrmions on the track. *Nat Nanotechnol* **8**, 152-156 (2013).
16. Krause S, Wiesendanger R. Skyrmionics gets hot. *Nat Mater* **15**, 493-494 (2016).
17. Wiesendanger R. Nanoscale magnetic skyrmions in metallic films and multilayers: a new twist for spintronics. *Nat Rev Mater* **1**, 16044 (2016).
18. Takashima R, Ishizuka H, Balents L. Quantum skyrmions in two-dimensional chiral magnets. *Phys Rev B* **94**, 134415 (2016).
19. Ochoa H, Tserkovnyak Y. Quantum skyrmionics. *Int J Mod Phys B* **33**, 1930005 (2019).
20. Tai J-SB, Smalyukh II. Surface anchoring as a control parameter for stabilizing torons, skyrmions, twisted walls, fingers, and their hybrids in chiral nematics. *Phys Rev E* **101**, 042702 (2020).
21. Lavrentovich O. Topological defects in dispersed liquid crystals, or words and worlds around liquid crystal drops. *Liq Cryst* **24**, 117-125 (1998).
22. Koshibae W, Nagaosa N. Theory of antiskyrmions in magnets. *Nat Commun* **7**, 10542 (2016).
23. Spaldin NA, Fiebig M, Mostovoy M. The toroidal moment in condensed-matter physics and its relation to the magnetoelectric effect. *J Phys Condens Matter* **20**, 434203 (2008).
24. Tolédano P, *et al.* Primary ferrotoroidicity in antiferromagnets. *Phys Rev B* **92**, 094431 (2015).
25. Spaldin NA, Fechner M, Bousquet E, Balatsky A, Nordström L. Monopole-based formalism for the diagonal magnetoelectric response. *Phys Rev B* **88**, 094429 (2013).
26. Spaldin NA, Ramesh R. Advances in magnetoelectric multiferroics. *Nat Mater* **18**, 203-212 (2019).
27. Chen X, *et al.* First-principles experimental demonstration of ferroelectricity in a thermotropic nematic liquid crystal: Polar domains and striking electro-optics. *Proc Natl Acad Sci USA* **117**, 14021-14031 (2020).

28. Nishikawa H, *et al.* A Fluid Liquid-Crystal Material with Highly Polar Order. *Adv Mater* **29**, 1702354 (2017).
29. Mandle RJ, Cowling SJ, Goodby JW. A nematic to nematic transformation exhibited by a rod-like liquid crystal. *Phys Chem Chem Phys* **19**, 11429-11435 (2017).
30. Chen X, Korblova E, Glaser MA, Maclennan JE, Walba DM, Clark NA. Polar in-plane surface orientation of a ferroelectric nematic liquid crystal: Polar monodomains and twisted state electro-optics. *Proc Natl Acad Sci U S A* **118**, e2104092118 (2021).
31. Mandle RJ, Sebastian N, Martinez-Perdiguero J, Mertelj A. On the molecular origins of the ferroelectric splay nematic phase. *Nat Commun* **12**, 4962 (2021).
32. Cherifi-Hertel S, *et al.* Non-Ising and chiral ferroelectric domain walls revealed by nonlinear optical microscopy. *Nat Commun* **8**, 15768 (2017).
33. Eremin A, *et al.* Pattern-stabilized decorated polar liquid-crystal fibers. *Phys Rev Lett* **109**, 017801 (2012).
34. Kaneshiro J, Uesu Y, Fukui T. Visibility of inverted domain structures using the second harmonic generation microscope: Comparison of interference and non-interference cases. *J Opt Soc Am B* **27**, 888-894 (2010).
35. Miyajima D, *et al.* Ferroelectric Columnar Liquid Crystal Featuring Confined Polar Groups Within Core-Shell Architecture. *Science* **336**, 209-213 (2012).
36. Zhao X, *et al.* Spontaneous helielectric nematic liquid crystals: Electric analog to helimagnets. *Proc Natl Acad Sci USA* **118**, e2111101118 (2021).
37. Brown S, *et al.* Multiple Polar and Non-polar Nematic Phases. *Chemphyschem* **22**, 2506-2510 (2021).
38. Zhao X, *et al.* Nontrivial phase matching in helielectric polarization helices: Universal phase matching theory, validation, and electric switching. *Proc Natl Acad Sci U S A* **119**, e2205636119 (2022).
39. Folcia CL, Ortega J, Vidal R, Sierra T, Etxebarria J. The ferroelectric nematic phase: an optimum liquid crystal candidate for nonlinear optics. *Liq Cryst* **49**, 899-906 (2022).
40. Ackerman PJ, Smalyukh II. Static three-dimensional topological solitons in fluid chiral ferromagnets and colloids. *Nat Mater* **16**, 426-432 (2017).
41. Chaikin PM, Lubensky TC. *Principles of Condensed Matter Physics*. Cambridge University Press (1995).
42. De Gennes P-G, Prost J. *The physics of liquid crystals*. Oxford university press (1993).
43. Toulouse G, Kléman M. Principles of a classification of defects in ordered media. *J Physique Lett* **37**, 149-151 (1976).
44. Mermin ND. The topological theory of defects in ordered media. *Rev Mod Phys* **51**, 591-648 (1979).
45. Manton N, Sutcliffe P. *Topological Solitons*. Cambridge University Press (2004).
46. Volovik GE, Mineev VP. Investigation of singularities in superfluid He$^3$ in liquid crystals by the homotopic topology methods. In: *30 Years of the Landau Institute — Selected Papers*) (1996).
47. Mertelj A, *et al.* Splay Nematic Phase. *Phys Rev X* **8**, 041025 (2018).
48. Li J, *et al.* Development of ferroelectric nematic fluids with giant ε dielectricity and nonlinear optical properties. *Sci Adv* **7**, eabf5047 (2021).
49. Sebastian N, *et al.* Ferroelectric-Ferroelastic Phase Transition in a Nematic Liquid Crystal. *Phys Rev Lett* **124**, 037801 (2020).
50. Li J, *et al.* How Far Can We Push the Rigid Oligomers/Polymers toward Ferroelectric Nematic Liquid Crystals? *J Am Chem Soc* **143**, 17857-17861 (2021).
51. Connor PLM, Mandle RJ. Chemically induced splay nematic phase with micron scale periodicity. *Soft Matter* **16**, 324-329 (2020).
52. Posnjak G, Copar S, Musevic I. Points, skyrmions and torons in chiral nematic droplets. *Sci Rep* **6**, 26361 (2016).
53. Pogna EAA, *et al.* Ultrafast, All Optically Reconfigurable, Nonlinear Nanoantenna. *ACS Nano* **15**, 11150-11157 (2021).

54. Kogo R, Araoka F, Uchida Y, Tamura R, Ishikawa K, Takezoe H. Second Harmonic Generation in a Paramagnetic All-Organic Chiral Smectic Liquid Crystal. *Appl Phys Express* **3**, 041701 (2010).
55. Volovik G, Lavrentovich O. Topological dynamics of defects: Boojums in nematic drops. *J Exp Theor Phys* **85**, 1997-2010 (1983).
56. Williams RD. Two transitions in tangentially anchored nematic droplets. *J Phys A: Math Gen* **19**, 3211-3222 (1986).
57. Fernandez-Nieves A, Link DR, Marquez M, Weitz DA. Topological changes in bipolar nematic droplets under flow. *Phys Rev Lett* **98**, 087801 (2007).
58. Lavrentovich O, Sergan V. Parity-breaking phase transition in tangentially anchored nematic drops. *Il Nuovo Cimento D* **12**, 1219-1222 (1990).
59. Ohzono T, Katoh K, Wang C, Fukazawa A, Yamaguchi S, Fukuda JI. Uncovering different states of topological defects in schlieren textures of a nematic liquid crystal. *Sci Rep* **7**, 16814 (2017).
60. Drzaic PS. A case of mistaken identity: spontaneous formation of twisted bipolar droplets from achiral nematic materials. *Liq Cryst* **26**, 623-627 (1999).
61. Wu J, Ma H, Chen S, Zhou X, Zhang Z. Study on concentric configuration of nematic liquid crystal droplet by Landau-de Gennes theory. *Liquid Crystals* **47**, 1698-1707 (2020).
62. Caimi F, et al. Superscreening and polarization control in confined ferroelectric nematic liquids. *arXiv preprint arXiv:221000886*, (2022).
63. Luk'yanchuk I, Tikhonov Y, Razumnaya A, Vinokur VM. Hopfions emerge in ferroelectrics. *Nat Commun* **11**, 2433 (2020).
64. Ravnik M, Žumer S. Landau–de Gennes modelling of nematic liquid crystal colloids. *Liquid Crystals* **36**, 1201-1214 (2009).
65. Kleman M, Lavrentovich OD. Topological point defects in nematic liquid crystals. *Philosophical Magazine* **86**, 4117-4137 (2006).
66. Hung FR, Guzman O, Gettelfinger BT, Abbott NL, de Pablo JJ. Anisotropic nanoparticles immersed in a nematic liquid crystal: defect structures and potentials of mean force. *Phys Rev E Stat Nonlin Soft Matter Phys* **74**, 011711 (2006).

*Acknowledgements*

S.A. acknowledges the supports from the National Key Research and Development Program of China (No. 2022YFA1405000) , the International Science and Technology Cooperation Program of Guangdong province (No. 2022A0505050006), General Program of Guangdong Natural Science Foundation (No. 2022A1515011026), National Natural Science Foundation of China for Young Scientists of China (NSFC No. 11904106), and International (Regional) Cooperation and Exchange Project (NSFC No. 12050410231). M.H acknowledges the support from the National Natural Science Foundation of China (NSFC No. 52273292). S.A and M.H acknowledge the Recruitment Program of Guangdong (No.2016ZT06C322) and the 111 Project (No. B18023).

*Figures*

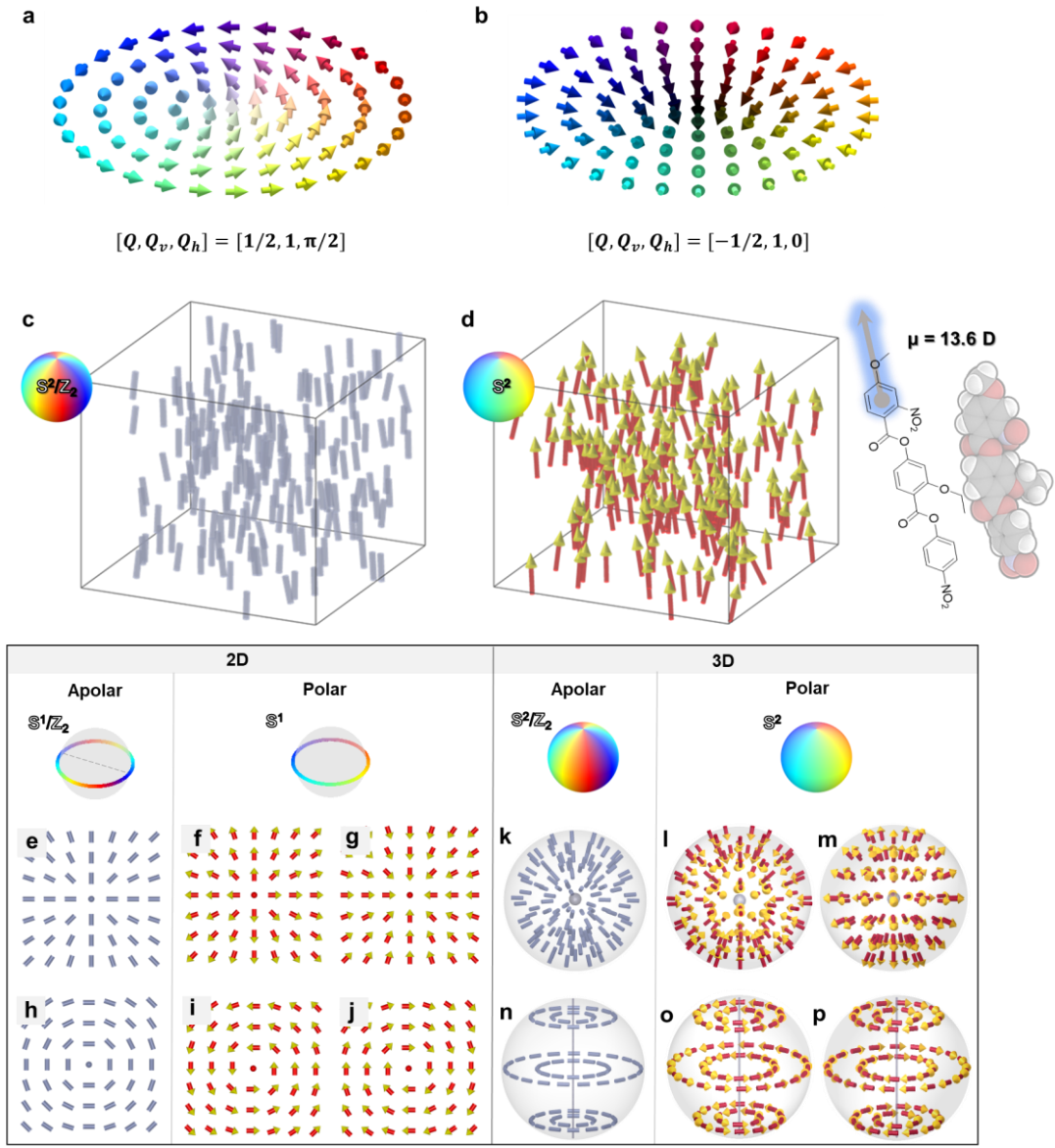

*Figure 1.* **Topological analogy between nonpolar and polar systems. a,b,** Illustrations of 2D merons with different sets of topological charge, vorticity number and helicity number, i.e. $[Q, Q_v, Q_h]$. (a) Bloch-type meron with $[Q, Q_v, Q_h] = [1/2, 1, \pi/2]$. (b) Neel-type anti-meron with $[Q, Q_v, Q_h] = [-1/2, 1, 0]$. **c**, Head-to-tail-equivalent director field of nonpolar nematics. **d**, Head-to-tail-inequivalent polarization field of ferroelectric nematics. The polarization entity of RM-OC$_2$ is shown. **e**, A +1 hedgehog-type defect in 2D nonpolar nematics. **f,g**, The polar counterparts of divergent (f) and convergent (g) polarization fields of (e). **h**, A concentric-type defect in 2D nonpolar nematics. **i,j**, The polar counterparts of right hand (i) and left hand (j) polarization fields of (h). **k**, A +1 hedgehog-type defect in 3D nonpolar nematics. **l,m**, The polar counterparts of divergent (l) and convergent (m) polarization fields of (k). **n**, A concentric-type defect in 3D nonpolar nematics. **o,p**, The trivial polar counterparts of right hand (o) and left hand (p) polarization fields of (n).

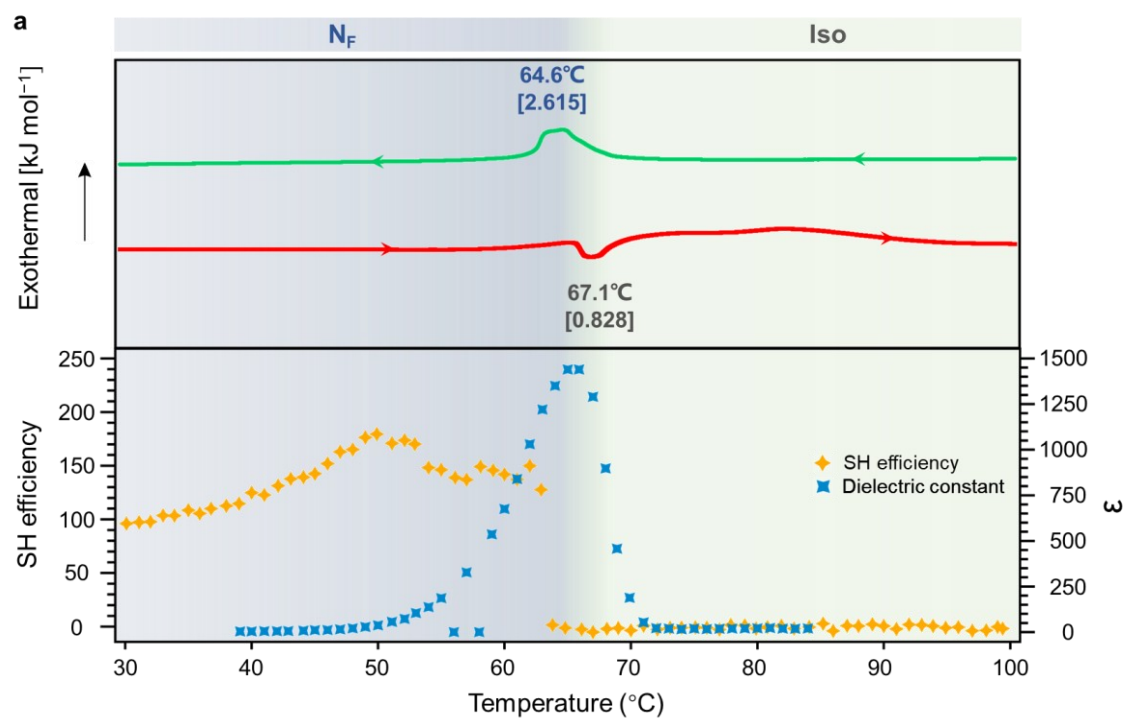

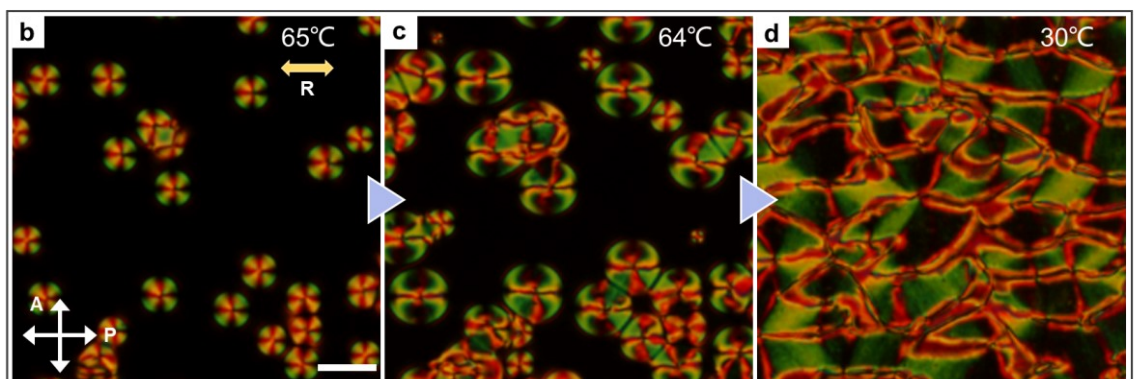

*Figure 2.* **General properties of the ferroelectric nematic. a,** Top chart shows DSC curves of RM-OC$_2$ during cooling (green line) and heating (red line) (rate: 3.0 K min$^{-1}$). Enthalpies of the corresponding phase transitions are shown in the square brackets [kJ mol$^{-1}$]. Bottom chart shows the temperature dependencies of the apparent dielectric constant at a frequency of 25 Hz (blue squares) and SH signal (yellow rhombuses). SH signal is defined as the SH signal intensity ratio of samples to that of a reference Y-cut quartz plate. **b-d,** The PLM texture evolution of RM-OC$_2$ during cooling. Optical graphs are taken within a planar cell. Rubbing direction is indicated by the yellow arrow. Scale bar: 50 μm.

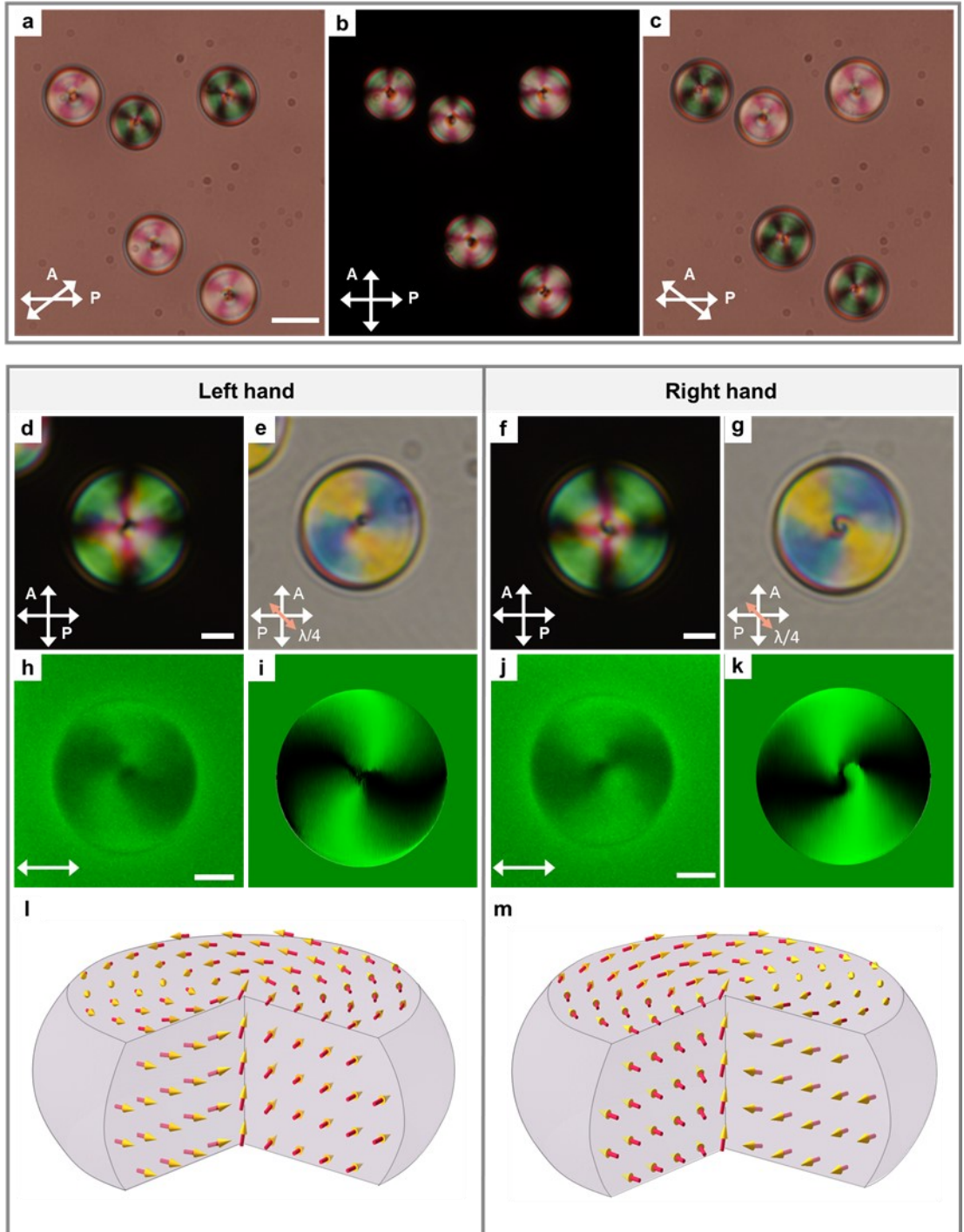

*Figure 3.* **N$_F$ droplets with opposite handedness. a-c**, PLM images of N$_F$ droplets under different polarizer conditions. The analyzer is rotated 52° clockwise (a); The analyzer is perpendicular to the polarizer (b); The analyzer is rotated 52° anticlockwise (c). Scale bar, 20 μm. **d-g**, PLM images of left hand (d-e) and right hand (f-g) N$_F$ droplets under crossed polarizers without (d, f), and with a quarter-wave plate (e, g). The angle between the fast axis of the quarter wave plate and the polarizer is 45°. The orange arrow indicates the fast axis of the quarter-wave plate. Scale bars, 5 μm. **h-k**, XY-cross-sectional FCPM images in the midplane of the cell visualized by a linearly polarization **P**. Scale bars, 5 μm. The white arrows represent the linear polarization as the incident light. The images of (h, j) were obtained experimentally and images of (i, k) are the numerical results. **l,m**, The calculation results were based on the reconstructed director fields obtained from FCPM data.

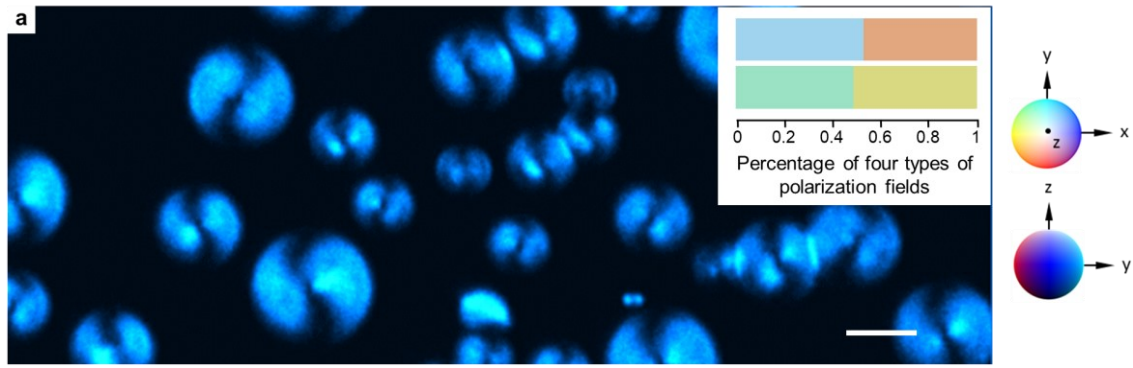

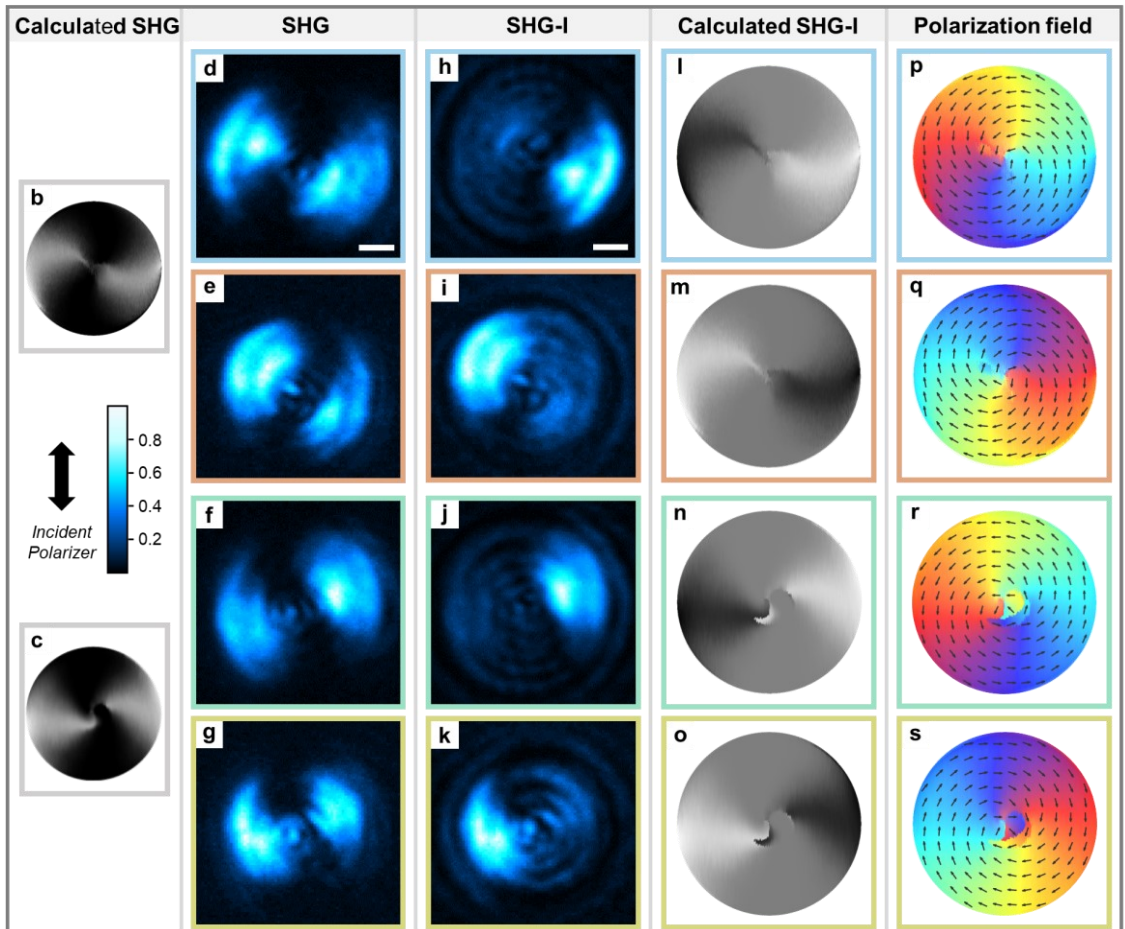

*Figure 4.* **Polarization textures probed by SHG and SHG-I microscopy a**, XY-cross-sectional SHG-CPM images of the $N_F$ droplets. The polarization of the incident laser is shown as the black double arrow. Scale bar, 10 μm. The inset shows the statistic number counts of four types of polarization fields observed in the experiment. **d-s**, Four types of the polarization fields **P(r)** are demonstrated in rectangles with blue, orange, green and yellow edges respectively. Experimental SHG microscopy (d-g) and SHG-I microscopy (h-k) observations of 4 types of polarization textures. In SHG microscopy observation, two types of polarization fields with the same chirality (d-e, f-g) are degenerate. In SHG-I microscopy observations (h-k), the interference conditions are the same. Scale bars, 5 μm. The signal reaches its maximum when polarization direction of the fundamental light is parallel to the electric polarization. **b,c**, Calculated SHG intensity obtained from fitted director fields **n(r)**. The director fields are obtained from FCPM data. **l-o**, Calculated SHG-I intensity obtain from fitted polarization fields **P(r)**. **p-s**, Reconstructed polarization fields with right-handed convergent polarization field (p), left-handed divergent polarization field (q), right-handed divergent polarization field (r) and left-handed convergent polarization field (s).

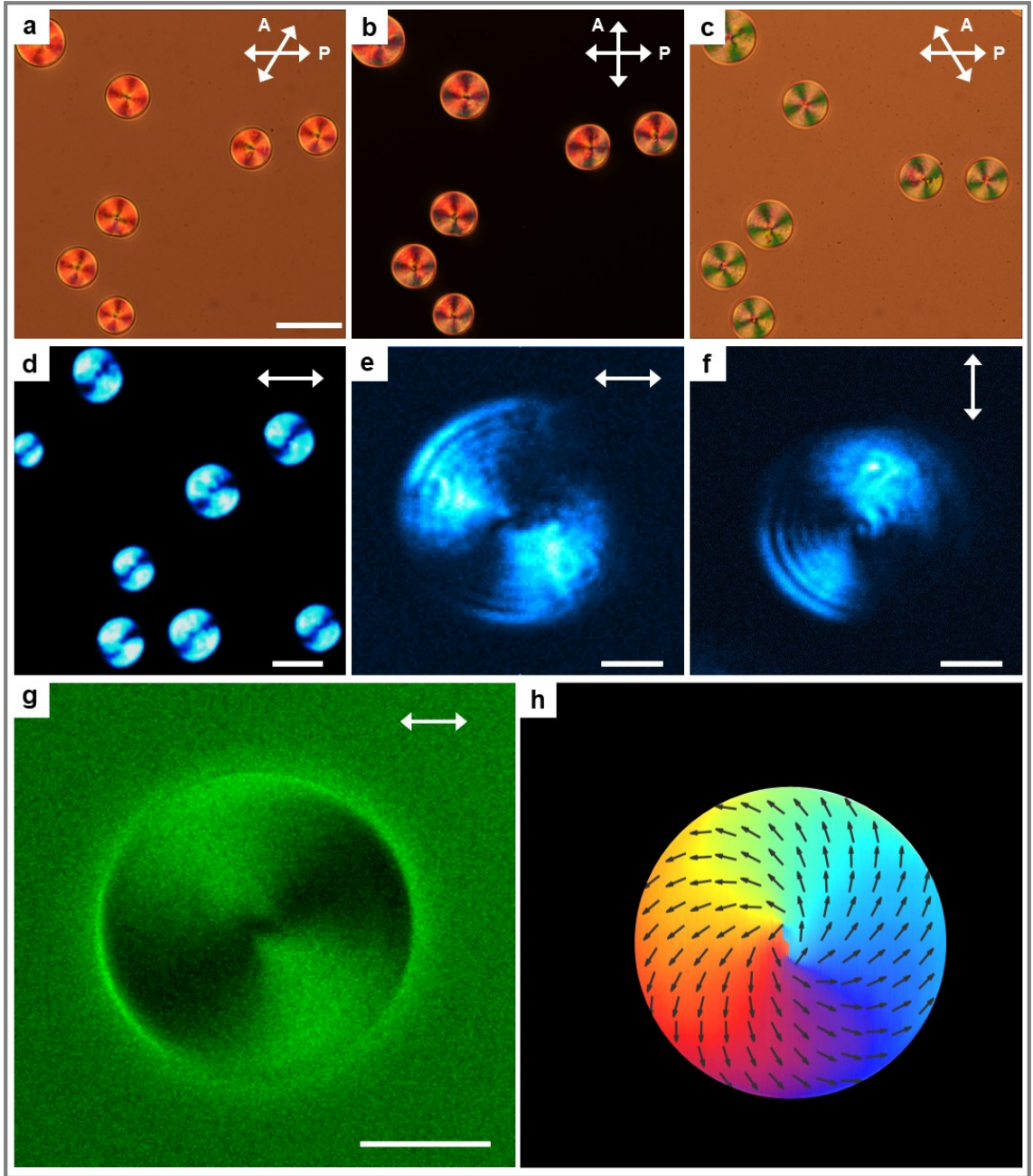

*Figure 5.* **Chirality biasing of N$_F$ droplets. a-c**, PLM texture of a RM-OC$_2$/R811 mixture (wt% = 99.5/0.5) taken under different combinations of polarizers. The analyzer is rotated 31°° clockwise (a) and 31° anticlockwise (c). Scale bar, 50 μm. **d**, XY-cross-sectional SHG-CPM image in the midplane of the cell visualized by a linearly polarization **P**. Scale bar, 20 μm. **e,f**, SHG microscopy images. Scale bars, 3 μm. **g**, XY-cross-sectional FCPM image visualized by a linearly polarization indicated by the white arrow. Scale bar, 10 μm. **h**, Reconstructed right-handed divergent polarization fields from SHG-I and FCPM data.

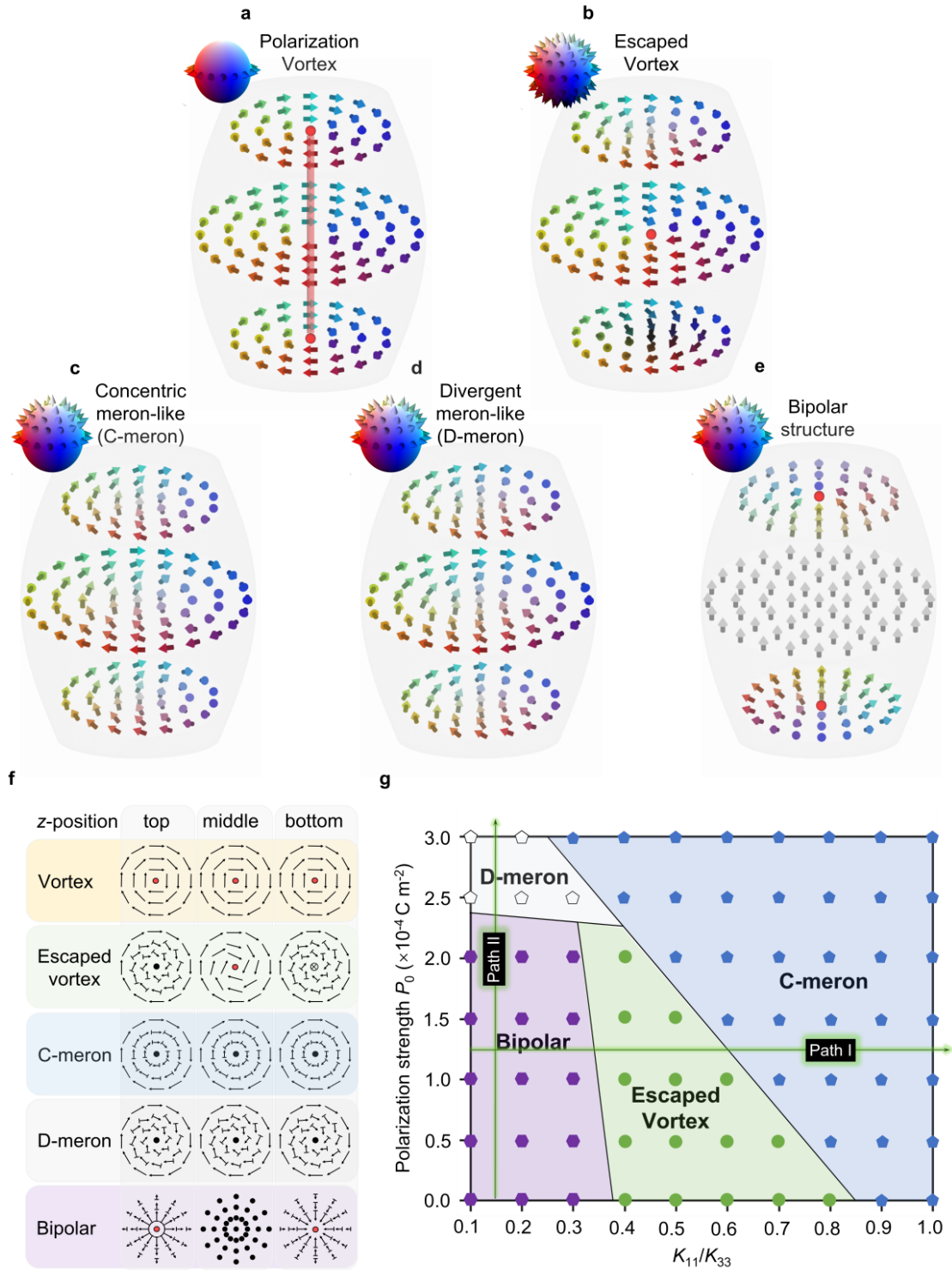

*Figure 6.* **Topological structures of the $N_F$ phase in cylindrically confined space. a-e**, Director fields and the corresponding order parameter space of simple polarization vortex (a), escaped vortex (b), concentric meron-like structure (c), divergent meron-like structure (d) and bipolar structure (e). **f**, The polarization fields projected on the xy-plane in different depth positions of the droplets are described nail vectors. Red points mean point or line defect. **g**, State diagram dependent on the ratio of the splay modulus to the bend modulus ($K_{11}/K_{33}$) and polarization strength $P_0$. The bipolar structure appears in the region where both $K_{11}$ and $P_0$ are small. Otherwise, it is replaced by the escaped vortex or D-meron structure.

# Supplementary Materials for
# Spontaneous electric-polarization topology in confined ferro-electric nematics

Jidan Yang[1]†, Yu Zou[1]†, Wentao Tang[1]†, Jinxing Li[1], Mingjun Huang[1,2]*, Satoshi Aya[1,2]*

Correspondence to: huangmj25@scut.edu.cn (M.H.); satoshiaya@scut.edu.cn (S.A.)

**This PDF file includes:**

Materials and Methods
Figures S1-S14
Supplementary Discussions 1-4

*Materials and Methods*

*Materials*

All commercial chemicals and solvents were used as received, unless stated otherwise. (R)-butan-2-ol, (S)-2-methylbutan-1-ol were obtained from Innochem. methyl 2-hydroxy-4-methoxybenzoate, 4-hydroxy-2-methoxybenzaldehyde, Sodium chlorite, Sodium dihydrogen phosphate were obtained from Energy Chemical. 4-nitrophenyl 4-hydroxybenzoate were obtained from Bidepharm. N-(3-dimethylaminopropyl)-N′-ethylcarbodiimide hydrochloride (EDC·HCl), and 4-dimethylaminopyridine (DMAP) were obtained from Sigma-Aldrich. Tetrahydrofuran (THF, Energy Chemical). Dichloromethane (DCM, Energy Chemical), Petroleum ether (PE, Energy Chemical) Ethyl acetate (EA, Energy Chemical), Methanol (MeOH, Energy Chemical, reagent grade), N,N-Dimethylformamide (DMF, Energy Chemical, reagent grade), Potassium carbonate, Potassium hydroxide were obtained from Sigma-Aldrich.

Characterizations of chemicals

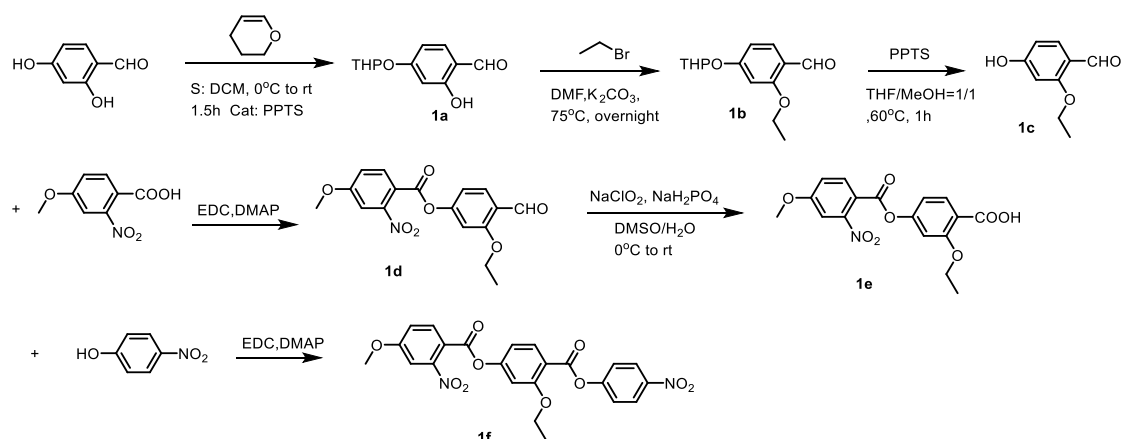

**2-hydroxy-4-((tetrahydro-2H-pyran-2-yl)oxy)benzaldehyde (1a):** A round bottom flask was charged with 2,4-dihydroxybenzaldehyde (2.6 g, 18.8 mmol), para-toluenesulfonic acid (0.196 g, 1.03 mol), and 100 mL diethyl ether forming a dispersion solution. The solution was degassed by bubbling nitrogen through for 5 min at ice bath. Then 3,4- Dihydro-2H-pyran (1.9 g, 2.06 mL, 22.6 mmol) was added dropwise via injector. The solution was slowly warmed to room temperature, and stirred under nitrogen for 5-6 h. The precipitate was then collected by filtration, washed with 50 mL diethyl ether (as little as possible), and dried in a vacuum oven. Yield: 2.3 g (55%); appearance: white powder. $^1$H NMR (500 MHz, Chloroform-*d*) δ 11.36 (s, 1H), 9.70 (s, 1H), 7.42 (d,

*J* = 8.6 Hz, 1H), 6.65 (dd, *J* = 8.6, 2.2 Hz, 1H), 6.61 (d, *J* = 2.2 Hz, 1H), 5.53 – 5.45 (m, 1H), 3.82 (td, *J* = 10.9, 10.5, 3.0 Hz, 1H), 3.66 – 3.60 (m, 1H), 1.86 (dt, *J* = 7.8, 3.8 Hz, 2H), 1.73 – 1.64 (m, 2H), 1.63 – 1.48 (m, 2H). $^{13}$C NMR (126 MHz, Chloroform-*d*) δ 194.67, 164.44, 164.24, 135.38, 115.82, 109.49, 103.74, 96.31, 62.25, 30.00, 25.01, 18.48.

**2-propoxy-4-((tetrahydro-2H-pyran-2-yl)oxy)benzaldehyde (1b):** A 100 mL round bottom flask was charged with the compound **1** (2.22 g, 10 mmol), potassium carbonate (4.15 g, 30 mmol), and 30 mL DMF. The solution was degassed by bubbling nitrogen for 3 min. Then bromoethane (0.9 mL, 1.31 g, 12 mmol) was added via injector. The solution was heated at 75°C and vigorously stirred for overnight. The mixture was cooled to room temperature, 200 mL water poured, and extracted with EA. The organic phase was washed with water and brine, then dried with anhydrous MgSO$_4$. The solvent was removed by rotary evaporation, and dry-loaded onto a silica gel column for purification using EA/hexane as eluent. Yield: 2.41 g, 96.3%, appearance: colorless oil. $^1$H NMR (400 MHz, Chloroform-*d*) δ 10.34 (s, 1H), 7.78 (d, *J* = 8.7 Hz, 1H), 6.75 – 6.53 (m, 2H), 5.51 (t, *J* = 3.1 Hz, 1H), 4.12 (qd, *J* = 7.0, 2.3 Hz, 2H), 3.93 – 3.79 (m, 1H), 3.70 – 3.56 (m, 1H), 2.06 – 1.93 (m, 1H), 1.87 (dt, *J* = 7.6, 3.6 Hz, 2H), 1.74 – 1.64 (m, 2H), 1.61 (dt, *J* = 12.1, 3.9 Hz, 1H), 1.47 (t, *J* = 7.0 Hz, 3H). $^{13}$C NMR (101 MHz, Chloroform-*d*) δ 188.60, 163.70, 163.11, 130.02, 119.42, 108.44, 100.35, 96.14, 64.15, 62.04, 30.06, 25.03, 18.40, 14.58.

**2-ethoxy-4-hydroxybenzaldehyde (1c):** A round bottom flask was charged with **1b** (2.0 g, 7.99 mmol), PPTS (2.01 g, 7.99mmol), 20 mL THF, and 20 mL ethanol. The solution was heated at 60 °C and monitored by TLC (6-48 h). The solvent was removed under reduced pressure. Water (40 mL) was added to the residue, and the product was extracted with DCM (3 × 50 mL). The organic phase was dried by anhydrous MgSO$_4$. The solvent was removed and the residue was purified by chromatography. The product was then concentrated and dried in the vacuum oven. Yield: 1.28 g (96.4%); appearance: white powder. $^1$H NMR (400 MHz, Chloroform-*d*) δ 10.32 (s, 1H), 7.77 (d, *J* = 8.4 Hz, 1H), 6.49 – 6.38 (m, 2H), 5.84 (s, 1H), 4.11 (q, *J* = 6.9 Hz, 2H), 1.47 (t, *J* = 7.0 Hz, 3H). $^{13}$C NMR (101 MHz, Chloroform-*d*) δ 188.51, 163.58, 162.70, 130.62, 119.01, 99.41, 64.21, 14.57.

**3-ethoxy-4-formylphenyl 4-methoxy-2-nitrobenzoate (1d):** A round bottom flask was charged with **1c** (1.2 g, 7.22 mmol), 4-methoxy-2-nitrobenzoic acid (1.57 g, 7.94 mmol), EDC (1.68 g, 10.83 mmol), N, N-dimethylaminopyridine (0.089 g, 0.722 mmol), then 50 mL dichloromethane solvent was added and cooled to 0 °C and flashed with nitrogen. The solution was stirred at 0 °C for 1 h then at room temperature for 17 hr. The solution was stripped of solvent by rotary evaporation, and dry-loaded onto a silica gel column for purification using dichloromethane/hexane as eluent. Yield: 1.95g (78.2%); appearance: white powder. $^1$H NMR (400 MHz, Chloroform-*d*) δ 10.45 (s, 1H), 7.91 (t, *J* = 8.6 Hz, 2H), 7.38 (d, *J* = 2.4 Hz, 1H), 7.21 (dd, *J* = 8.7, 2.5 Hz, 1H), 6.88 (p, *J* = 3.1, 2.6 Hz, 2H), 4.17 (q, *J* = 7.0 Hz, 2H), 3.96 (d, *J* = 1.6 Hz, 3H), 1.50 (t, *J* = 7.0 Hz, 3H). $^{13}$C NMR (101 MHz, Chloroform-*d*) δ 188.83, 162.94, 162.60, 162.39, 156.27, 132.48, 130.58, 129.73, 123.01, 117.88, 117.28, 113.70, 109.74, 106.02, 64.66, 56.34, 14.51.

**2-ethoxy-4-((4-methoxy-2-nitrobenzoyl)oxy)benzoic acid (1e):** Sodium dihydrogen phosphate (2.64 g, 22.01 mmol) and sodium chlorite (1.74 g, 19.26 mmol) were dissolved in water (30 mL) and slowly added to a stirred solution of compound **1d** (1.9 g, 5.5 mmol) in DMSO (40 mL) by constant pressure dropping funnel at 0 °C. The mixture was allowed to warm to room temperature and stir for 6 hours. The mixture was diluted with water and solid NaHCO$_3$ was added to adjust the pH of the solution to 9. The solution was washed with ethyl acetate, then the pH was adjusted to 4 by the addition of 1M HCl solution and extracted with ethyl acetate. The combined organic extracts ware washed with brine, dried (NaSO$_4$). The solvent was removed by rotary evaporation, and dry-loaded onto a silica gel column for purification using EA/hexane as eluent. Yield: 1.92 g, 96.6%, appearance: white powder. $^1$H NMR (500 MHz, DMSO-$d_6$) δ 12.60 (s, 1H), 8.09 (d, $J$ = 8.7 Hz, 1H), 7.70 (d, $J$ = 8.4 Hz, 1H), 7.64 (d, $J$ = 2.5 Hz, 1H), 7.38 (dd, $J$ = 8.7, 2.5 Hz, 1H), 7.01 (d, $J$ = 2.1 Hz, 1H), 6.83 (dd, $J$ = 8.4, 2.1 Hz, 1H), 4.05 (q, $J$ = 6.9 Hz, 2H), 3.91 (s, 3H), 1.29 (t, $J$ = 6.9 Hz, 3H). $^{13}$C NMR (126 MHz, DMSO-$d_6$) δ 167.19, 163.68, 162.37, 159.06, 153.91, 151.53, 133.45, 132.39, 120.25, 118.49, 115.51, 113.57, 110.39, 107.77, 65.01, 57.23, 14.96.

**3-ethoxy-4-((4-nitrophenoxy)carbonyl)phenyl 4-methoxy-2-nitrobenzoate (1f)**: A round bottom flask was charged with **1e** (1.8 g, 4.98 mmol), 4-nitrophenol (0.76 g, 5.48 mmol), EDC (1.16 g, 7.47 mmol), N, N-dimethylaminopyridine (36.5 mg, 0.3 mmol), then 50 mL dichloromethane solvent was added and cooled to 0 °C and flashed with nitrogen. The solution was stirred at 0 °C for 1 h then at room temperature for 17 hr. The solution was stripped of solvent by rotary evaporation, and dry-loaded onto a silica gel column for purification using dichloromethane/hexane as eluent. Yield: 1.86g (77.4%); appearance: white powder. $^1$H NMR (500 MHz, Chloroform-$d$) δ 8.35 – 8.24 (m, 2H), 8.12 – 8.04 (m, 1H), 7.92 (d, $J$ = 8.6 Hz, 1H), 7.45 – 7.34 (m, 3H), 7.20 (dd, $J$ = 8.6, 2.5 Hz, 1H), 6.97 – 6.87 (m, 2H), 4.17 (q, $J$ = 7.0 Hz, 2H), 3.95 (s, 3H), 1.48 (t, $J$ = 7.0 Hz, 3H). $^{13}$C NMR (126 MHz, Chloroform-$d$) δ 163.07, 162.69 (d, $J$ = 3.7 Hz), 161.04, 162.70, 162.67, 155.94, 155.64, 150.87, 145.38, 133.78, 132.57, 125.30, 122.77, 117.97, 117.31, 116.02, 113.28, 109.84, 106.85, 65.15, 56.42, 14.63.

*Figures S1-S13*

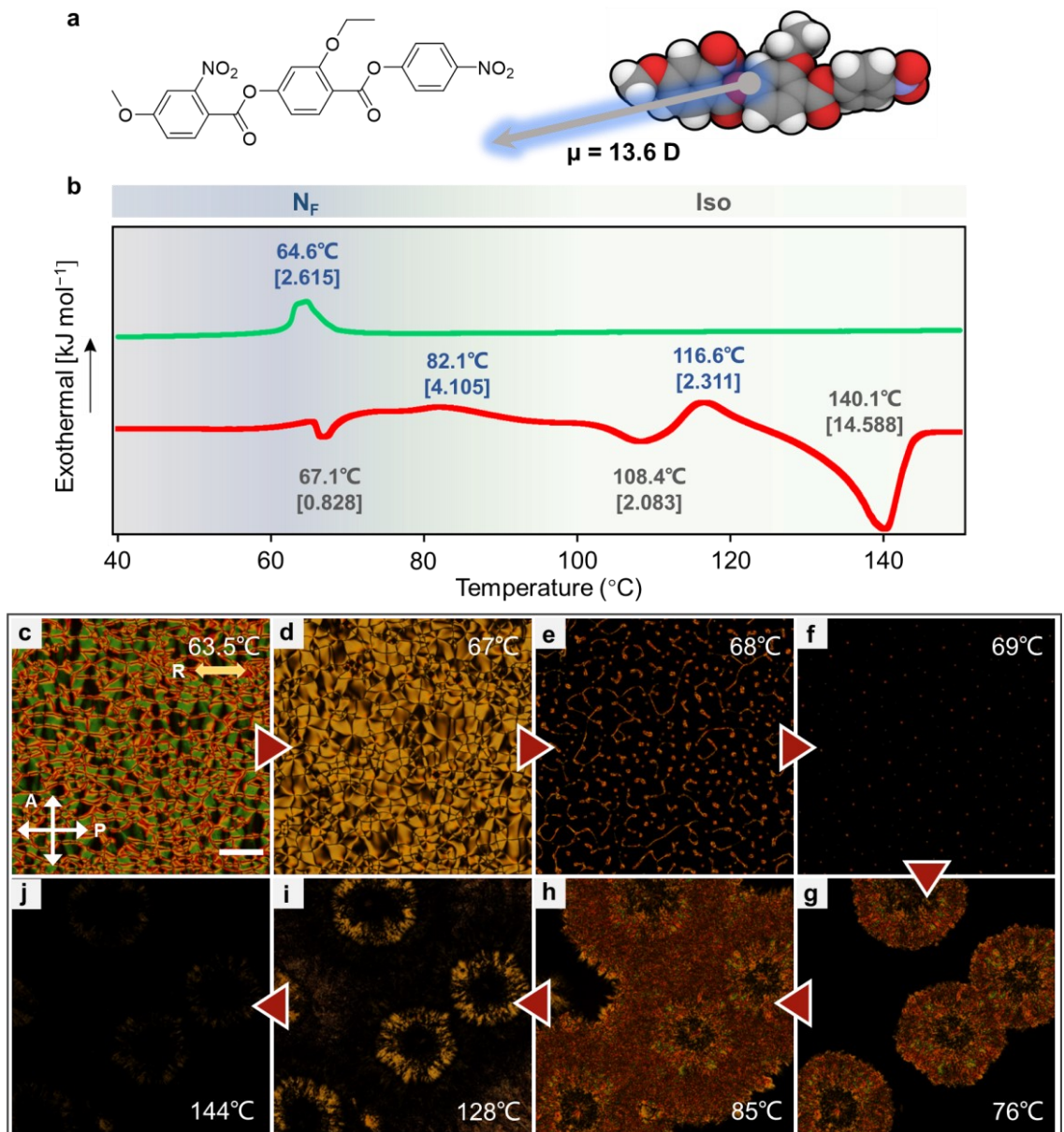

Figure **S1.** DSC curves and PLM images of the ferroelectric nematic. **a**, Chemical structure of RM-OC$_2$. **b**, DSC curves during cooling (green line) and heating (red line) at a scan rate of 3.0 K min$^{-1}$. Enthalpies of corresponding phase transitions are shown in the square brackets in the unit of kJ mol$^{-1}$. **c-j**, The PLM texture evolution of RM-OC$_2$ during heating. During the heating process, the system undergoes the crystallization after going into the Iso phase. Optical graphs are taken in an antiparallelly rubbed cell. Rubbing direction is indicated by the yellow arrow. Scale bar: 100 μm.

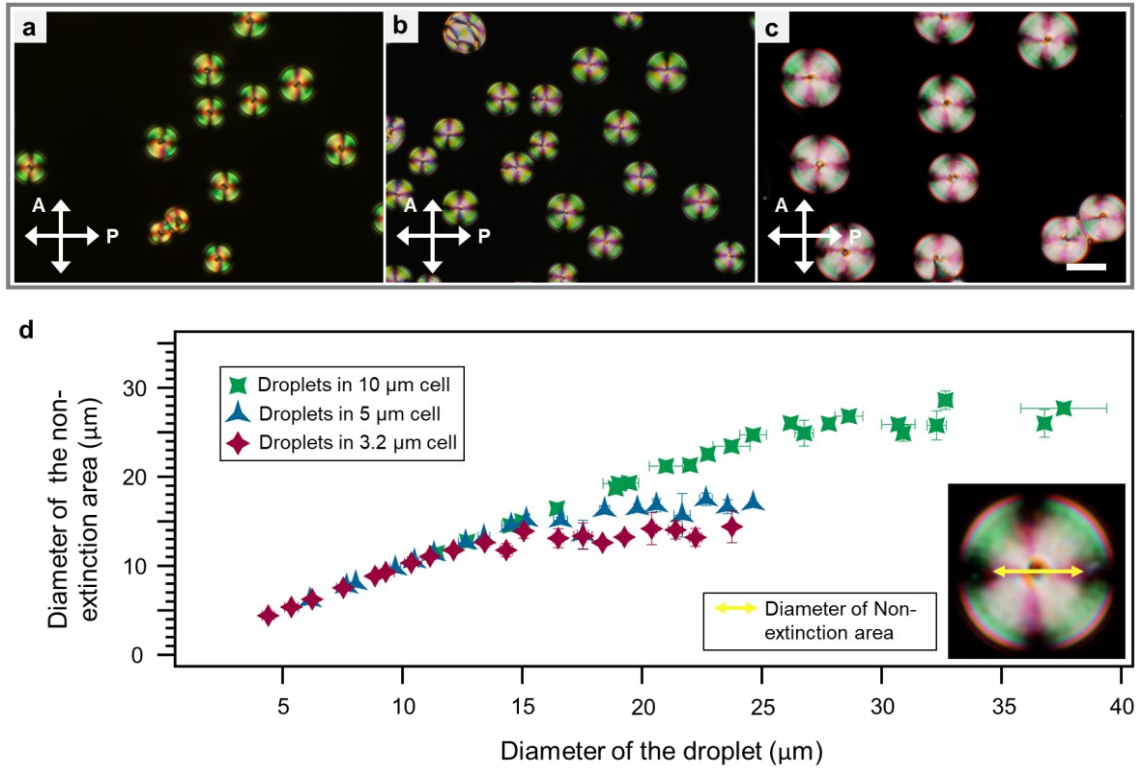

Figure **S2.** PLM characterization of $N_F$ droplets. **a-c**, PLM images of the $N_F$ droplets under crossed polarizers in homemade LC cells with the thickness of 3.2 μm (a), 5 μm (b), and 10 μm (c). Scale bar: 20 μm. **d**, The diagram showing the relationship between the non-extinction area in droplet center (indicated by a yellow arrow) and the diameter of the emerging $N_F$ droplets in different film thicknesses.

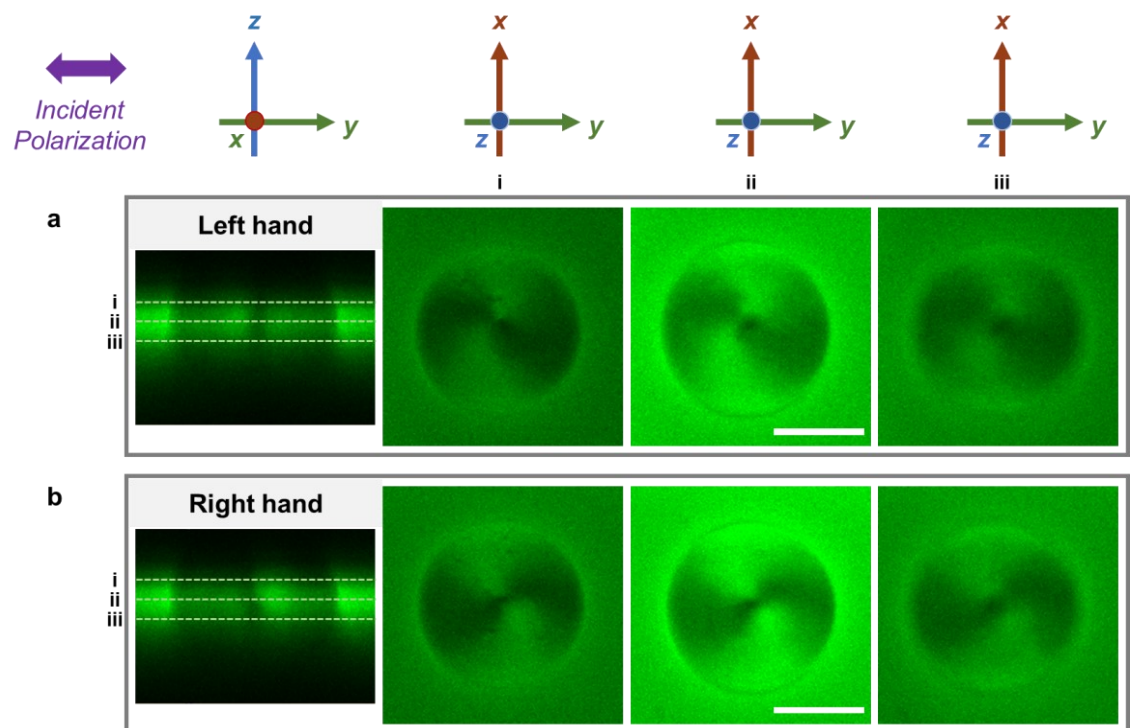

Figure **S3.** Fluorescence confocal polarizing microscopy (FCPM) images of $N_F$ droplets with the opposite handedness. The purple arrow represents the linear polarization of the incident light. **a,b**, Cross-sectional FCPM images of the lefthand $N_F$ droplet (a) and the right hand $N_F$ droplet (b). XY- and YZ-cross-sectional images are visualized by a linearly polarization at three different positions. Scale bars, 10 μm.

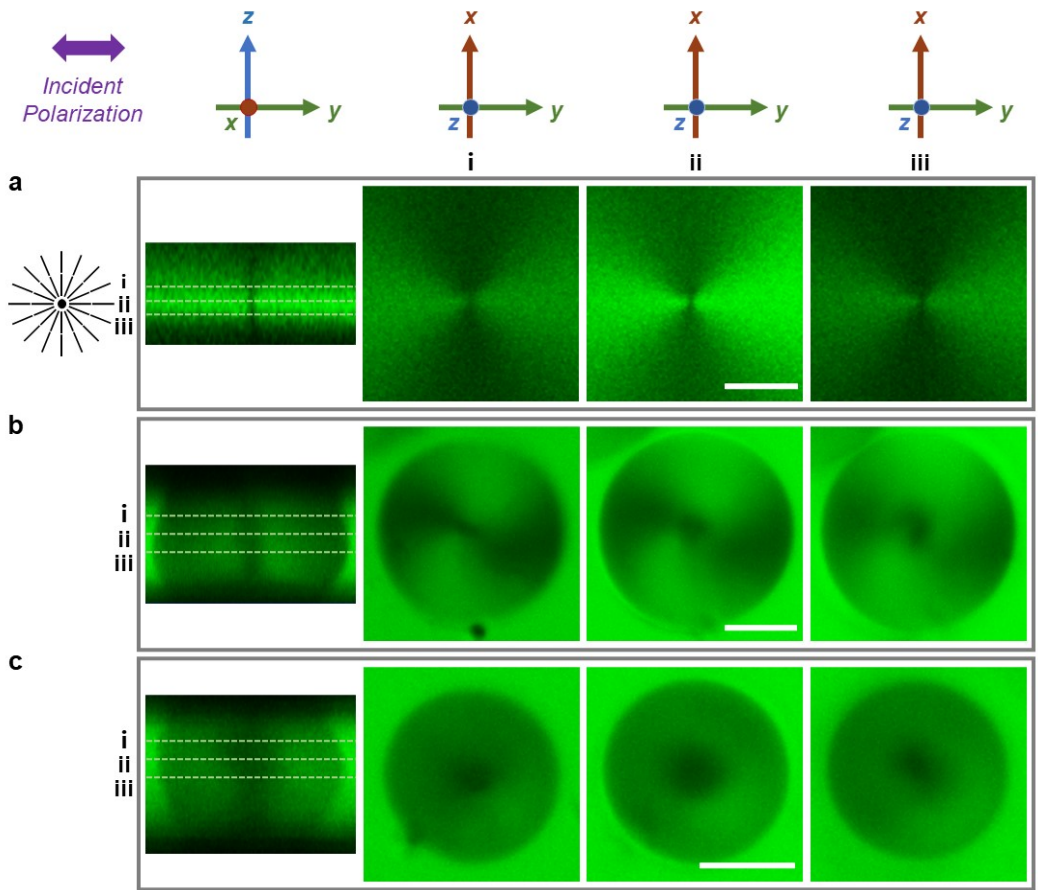

Figure **S4.** FCPM images for a photoaligned pattern and NF droplets. The purple arrow represents the linear polarization of the incident light. **a**, XY- and YZ-cross-sectional FCPM images of a +1 disclination line that connects the top and bottom substrates in a 5 μm cell created by the photoalignment technique. We use a commercial 5CB doped with Nile Red (c = 0.05 %) to observe the line disclination under FCPM. The XY-cross-sectional images are visualized by the linearly polarization at three different positions, corresponding to i~iii. Scale bar, 40 μm. An objective lens with the numerical aperture NA=0.8 is used for the observations (Plan-ApoCHROMAT20x, Zeiss). **b**, XY-cross-sectional FCPM images of an $N_F$ droplet of DIO (see Supplementary Discussion 2). The XY-cross-sectional images are visualized by a linearly polarization at three different positions, corresponding to i~iii in the YZ-cross-sections. Scale bar, 5 μm. An oil-immersion objective lens with the numerical aperture NA=1.4 is used for the observations (Plan-ApoCHROMAT63x, oil, Zeiss). A 5-μm cell with a planar alignment without rubbing is used for observation. **c**, XY-cross-sectional FCPM images of an $N_F$ droplet of DIO. The XY-cross-sectional images are visualized by a circular polarization at three different positions, corresponding to i~iii. Scale bar, 5 μm.

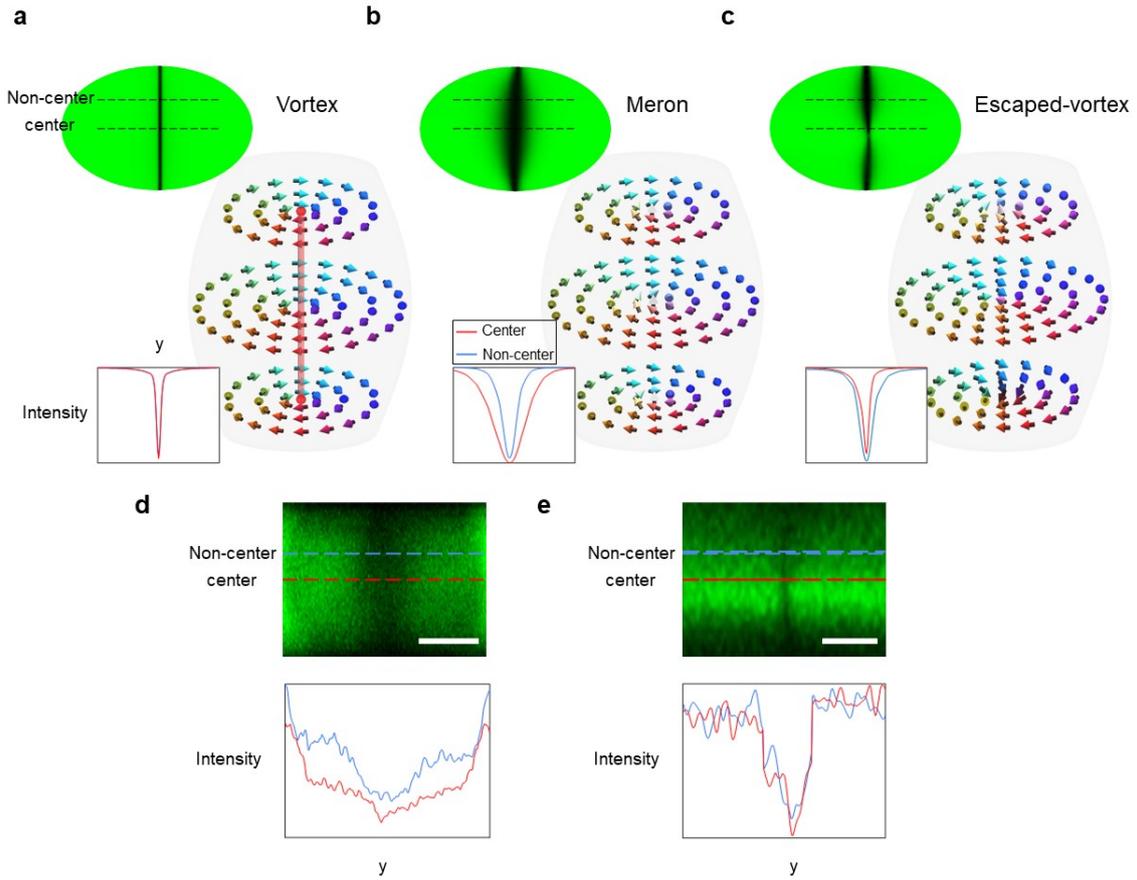

Figure **S5.** Fluorescence intensity profiles. **a-c**, Simulated structures of vortex (a), meron (b) and escaped vortex (c). Calculated YZ-cross-sections of FCPM images and signal profiles obtained from the simulated structures. The simulated FCPM images are visualized by a circular polarization. **d**,**e**, Experimental FCPM profiles of a $N_F$ droplet (d) and the disclination line (e). The scale bar in (d) is 3 µm and in (e) is 20 µm. The experimental FCPM images (d) are visualized by a circular polarization and (e) are visualized by a linearly polarization. The red lines represent the intensity profile along the path in the center area of the droplet. The blue lines represent the intensity profile along the path in the non-center area of the droplet.

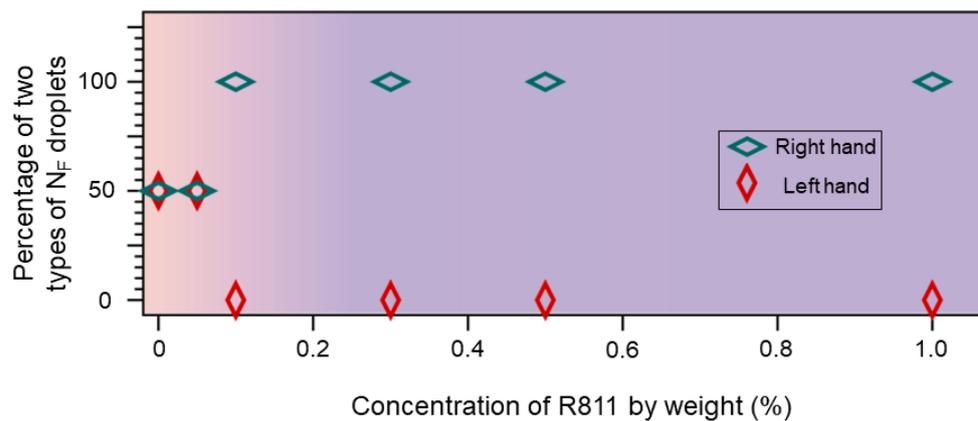

Figure **S6.** $N_F$ droplet statistics. Percentage of two types of $N_F$ droplets with the opposite handedness for RM-OC$_2$/R811 mixtures as a function of the weight percentage of R811.

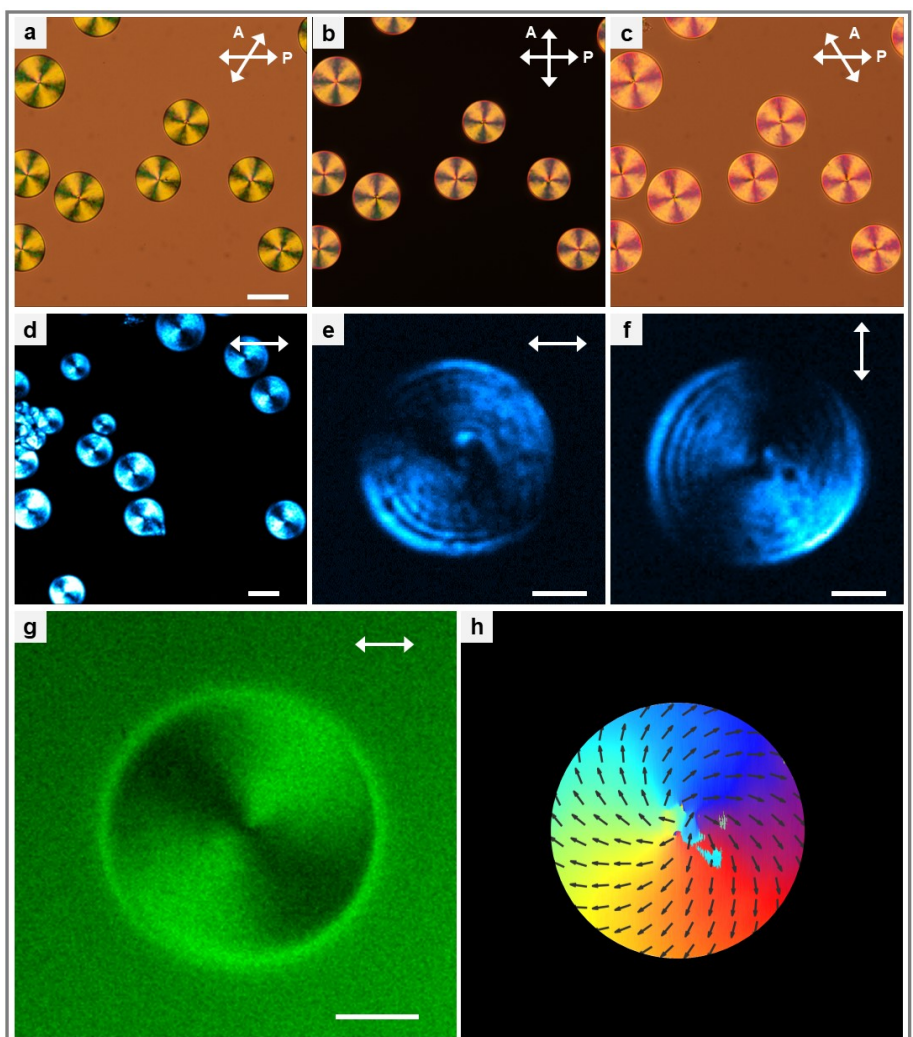

Figure **S7.** Chirality biasing of $N_F$ droplets. **a-c**, PLM textures of an RM-OC$_2$/S811 mixture (99.5/0.5 in wt%) taken under different combinations of polarizers. The analyzer is rotated 32° clockwise (a) and anticlockwise (c). Scale bar, 50 μm. **d**, XY-cross-sectional SHG-CPM image, Scale bar, 20 μm. **e**,**f**, SHG microscopy images with two different incident laser polarizations, Scale bars, 3 μm. **g**, XY-cross-sectional FCPM image visualized by a linearly polarization indicated by the white arrow, Scale bar, 10 μm. **h**, Reconstructed left-handed divergent polarization fields from SHG-I and FCPM data.

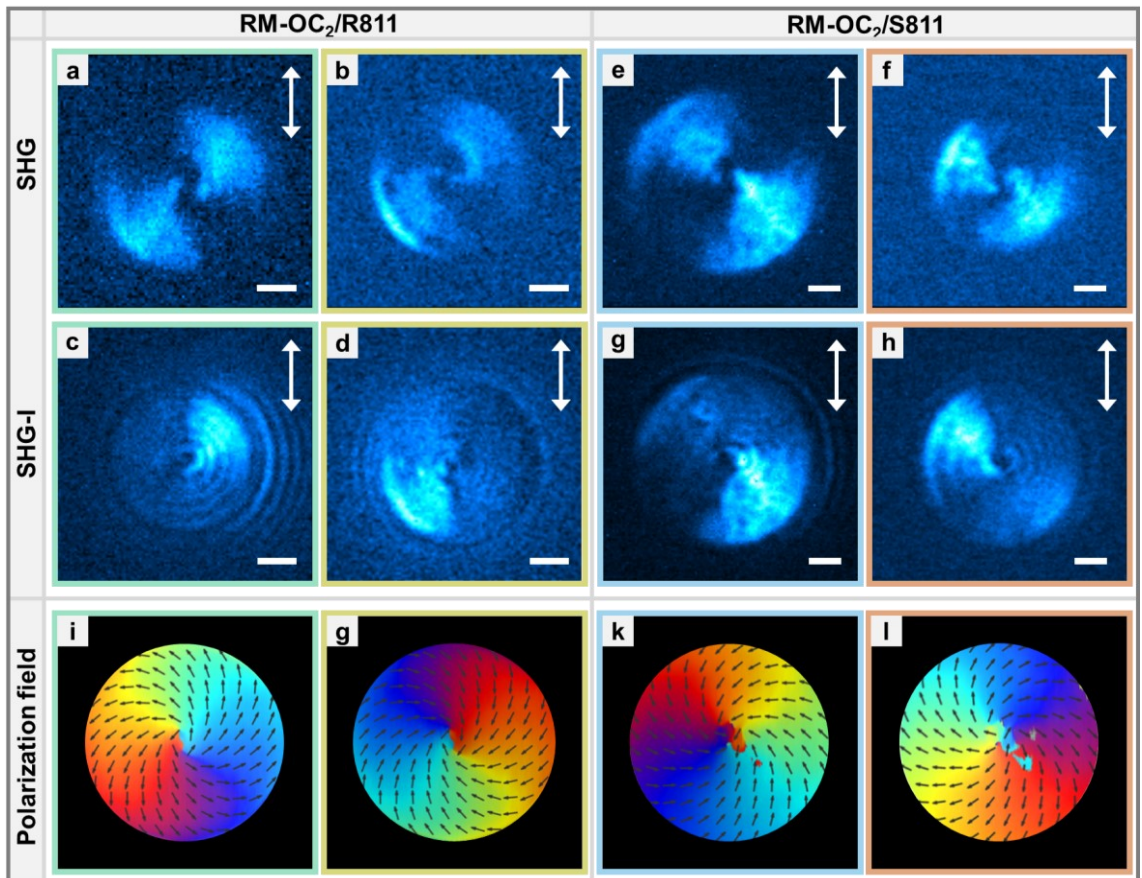

Figure **S8.** Polarization textures of chirality biased droplets probed by SHG and SHG-I microscopy. **a-d**, Experimental SHG microscopy (a, b) and SHG-I microscopy (c, d) observations of polarization textures of RM-OC$_2$/R811 mixture (99.5/0.5 in wt%). **e-h**, Experimental SHG microscopy (e, f) and SHG-I microscopy (g, h) observations of polarization textures of RM-OC$_2$/S811 mixture (99.5/0.5 in wt%). In SHG-I microscopy observations (c, d, g, h), the interference conditions are the same. Scale bars, 5 μm. **i-l**, Reconstructed polarization fields with right-handed divergent polarization field (i), left-handed convergent polarization field (g), right-handed convergent polarization field (k) and left-handed divergent polarization field (l).

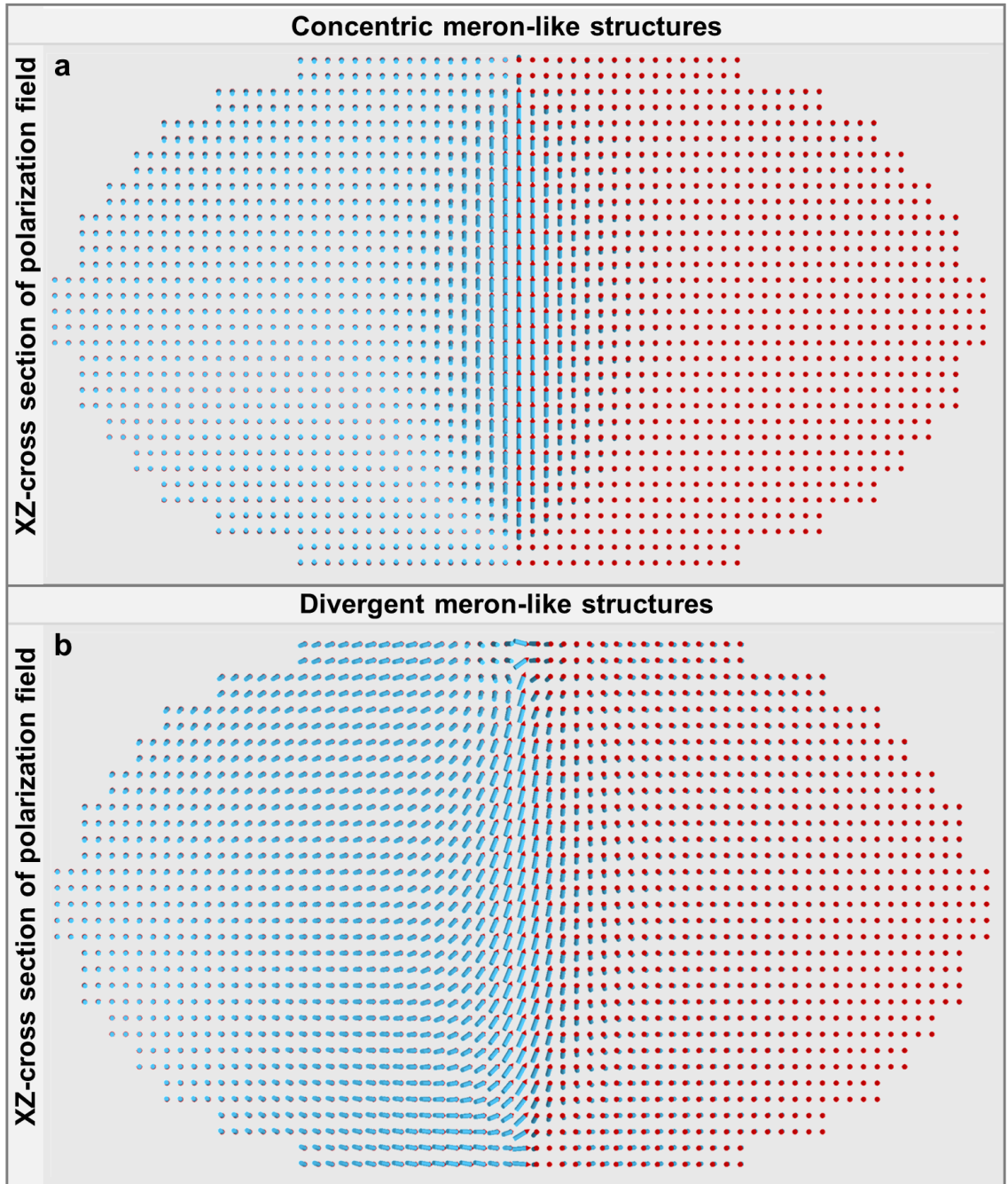

Figure **S9.** Comparison of simulation results between concentric meron-like structure and divergent meron-like structure in confined space. **a,b**, The XZ-cross-sectional polarization field of concentric meron-like structure (a) and divergent meron-like structure (b).

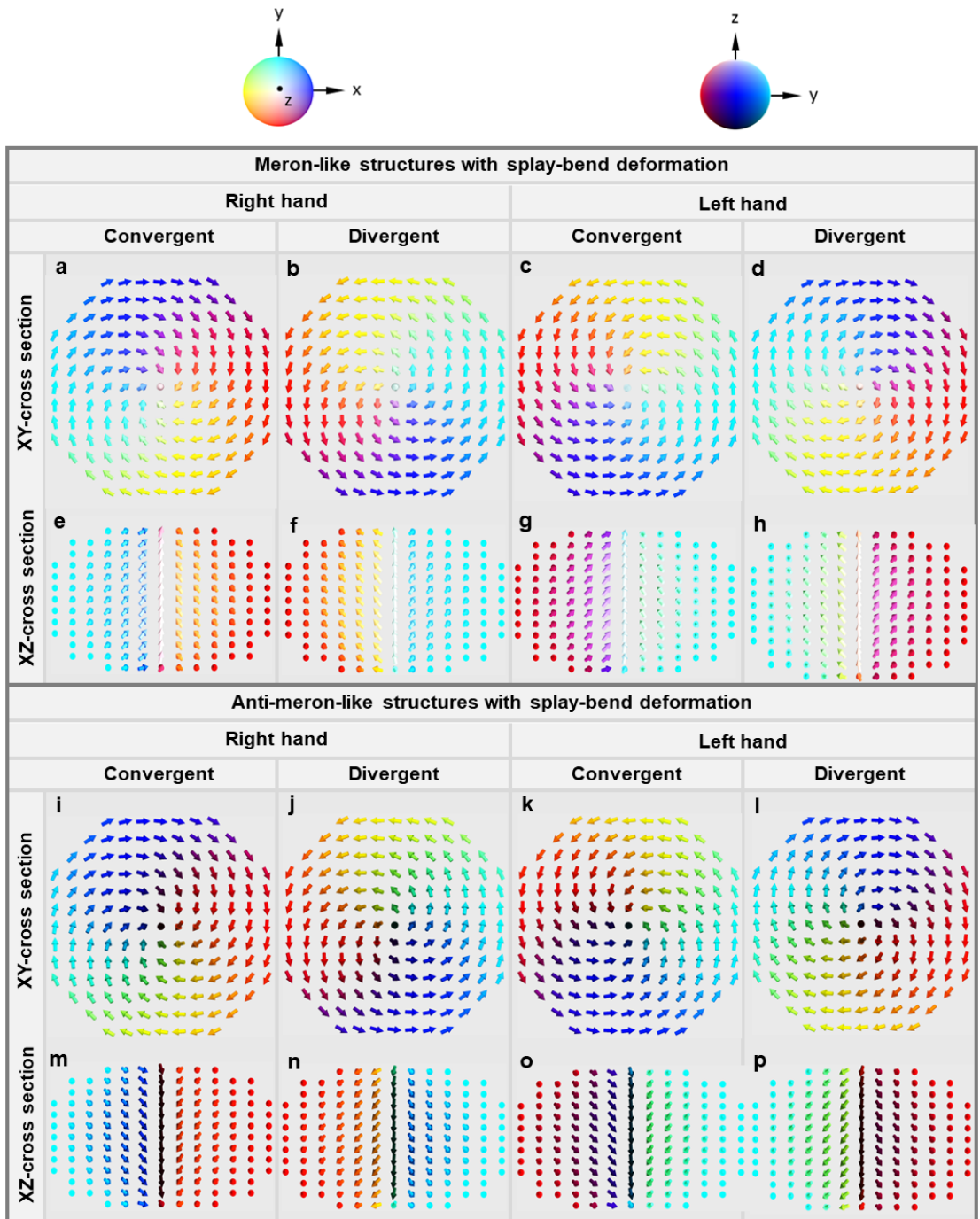

Figure **S10.** Numerical simulation results of possible divergent meron-like structure states and their corresponding anti-meron-like structures in confined space. **a-h**, Four types of polarization fields of divergent meron-like structure classified by the combination of handedness of the spirals and polarization orientation. **i-p**, Four types of polarization fields of the divergent (convergent) anti-meron-like structures classified by the combination of their handedness of the spirals handedness and polarization orientation.

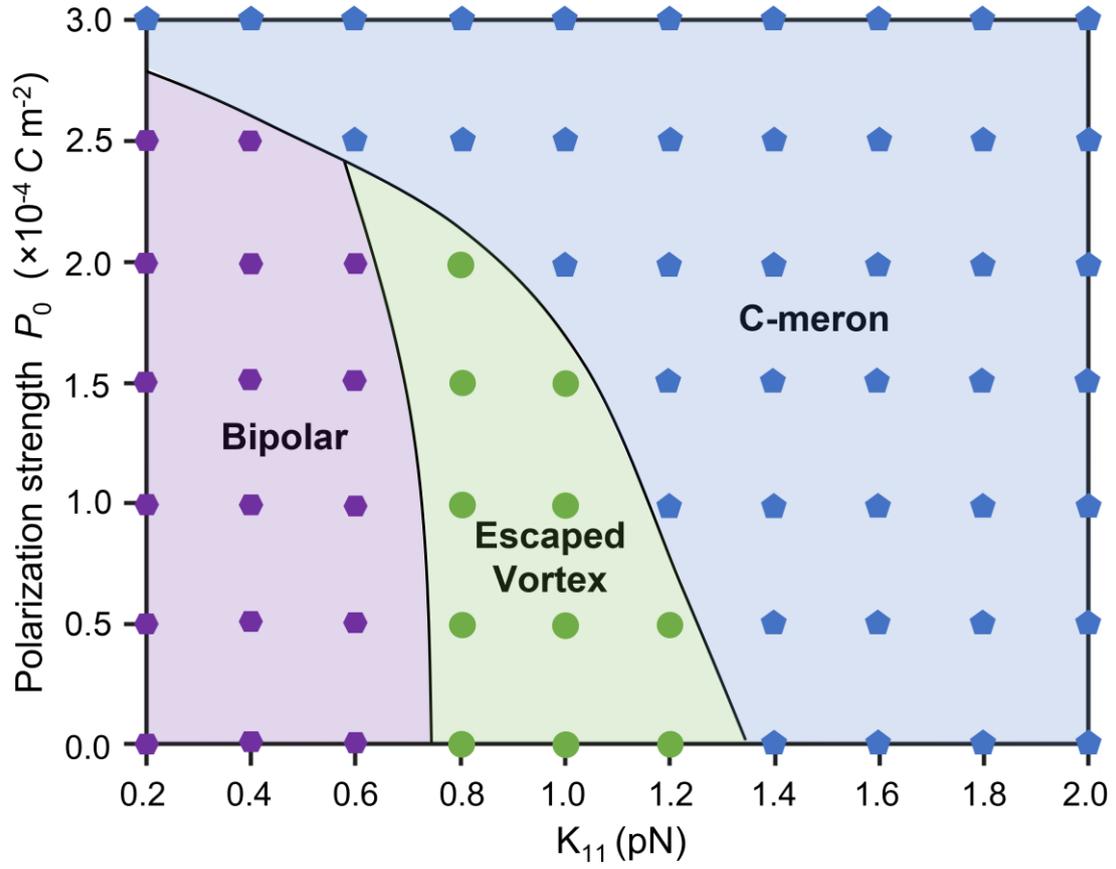

**Figure S11.** State diagram dependent on the splay modulus $K_{11}$ and polarization strength $P_0$. We assume that the twist elastic modulus is equal to the bend elastic modulus, $K_{22} = K_{33} = 2$ pN. In the range of $P_0 < 2.5 \times 10^{-4}$ C m$^{-2}$, as $K_{11}$ decreases, the structure with the lowest energy changes from C-meron to escaped vortex structure, and then to bipolar structure. The reason for these transformations is that the above three structures contain more splay deformation in turn. However, in the region of larger polarity, $P_0 > 2.5 \times 10^{-4}$ C m$^{-2}$, the bipolar structure with two +1 point defects at its two poles will consume huge energy due to the sharp increase of dipolar interactions, so C-meron is again the most favorable even in the range of small $K_{11}$.

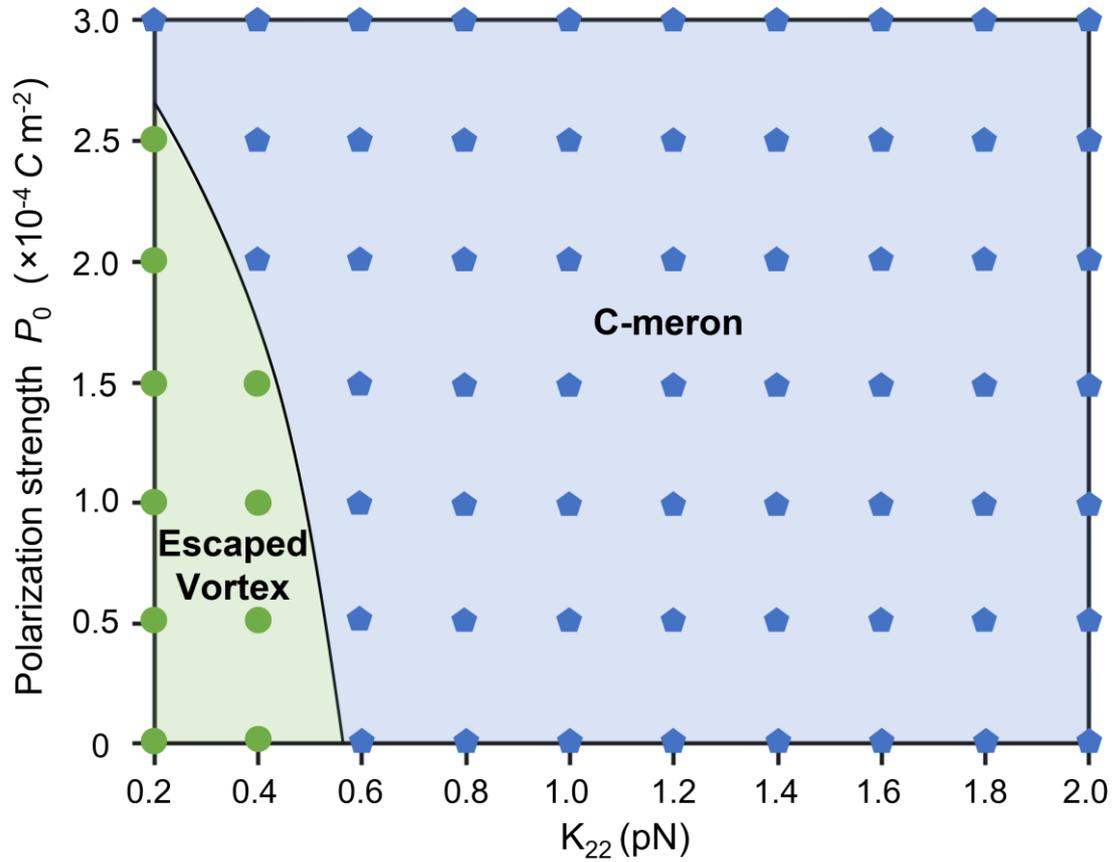

Figure S12. State diagram dependent on the twist modulus $K_{22}$ and polarization strength $P_0$. We assume that the splay elastic modulus is equal to the bend elastic modulus, $K_{11} = K_{33} = 2$ pN. In the range of $P_0 < 3.0 \times 10^{-4}$ C m$^{-2}$, as $K_{22}$ decreases, the structure with the lowest energy changes from C-meron to escaped vortex structure, since the very small twist elastic modulus $K_{22}$ breaks the symmetry of the system and thus spontaneously forms chiral structures in the non-chiral system. However, this effect is suppressed in the high polarity region, i.e. $P_0 = 3.0 \times 10^{-4}$ C m$^{-2}$, since the escaped vortex structure contains large energy from dipolar interaction near the -1 point defect in its center.

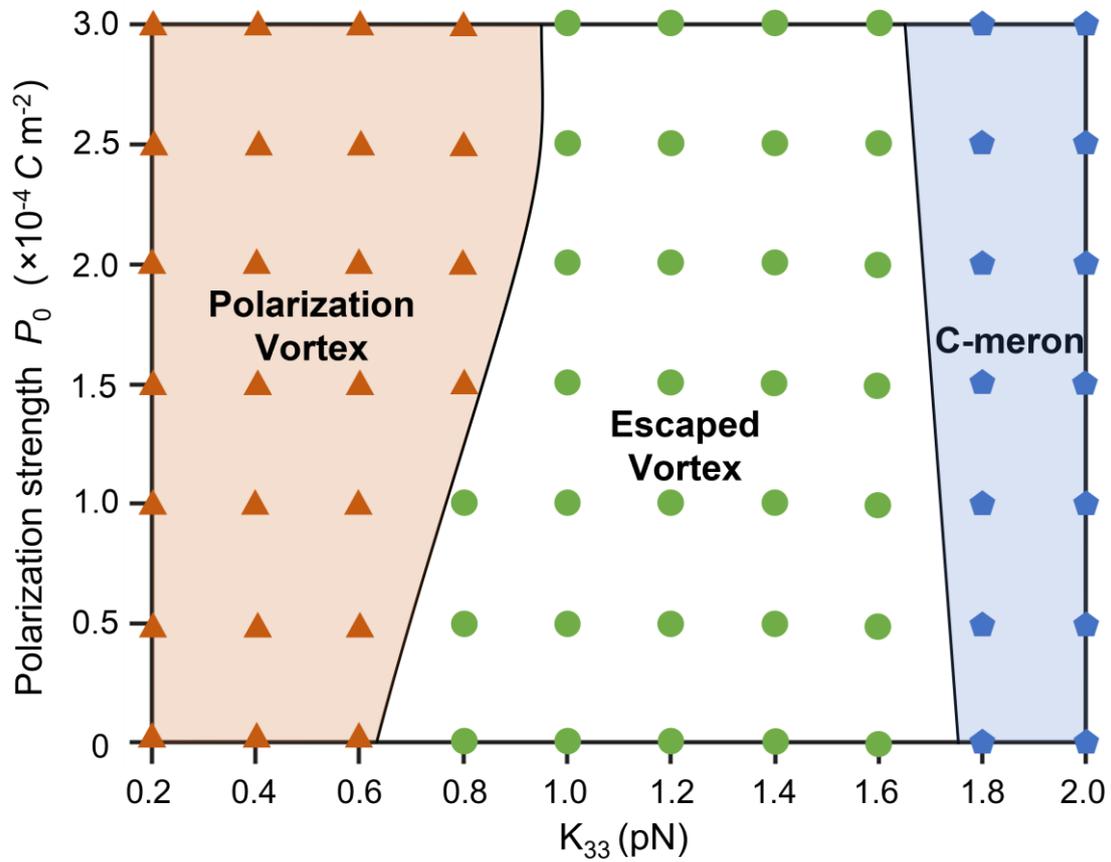

Figure **S13**. State diagram dependent on the bend modulus $K_{33}$ and polarization strength $P_0$. We assume that the splay elastic modulus is equal to the twist elastic modulus, $K_{11} = K_{22} = 2$ pN. As the bend modulus $K_{33}$ decreases, the structure with the lowest energy changes from C-meron to escaped vortex structure, and then to polarization vortex structure. The escaped vortex structure here is evolved from the polarization vortex structure which has a lot of bend deformation and has a defect line in its core. In order to reduce the total energy of structure, the LC molecules near the polarization vortex structure's core will escape along the z direction so that the line defect will degenerate into a point defect. The polarity affects the phase area of the three structures (high polarity shrinks the phase diagram area of the escaped vortex structure and increases that of the other two structures).

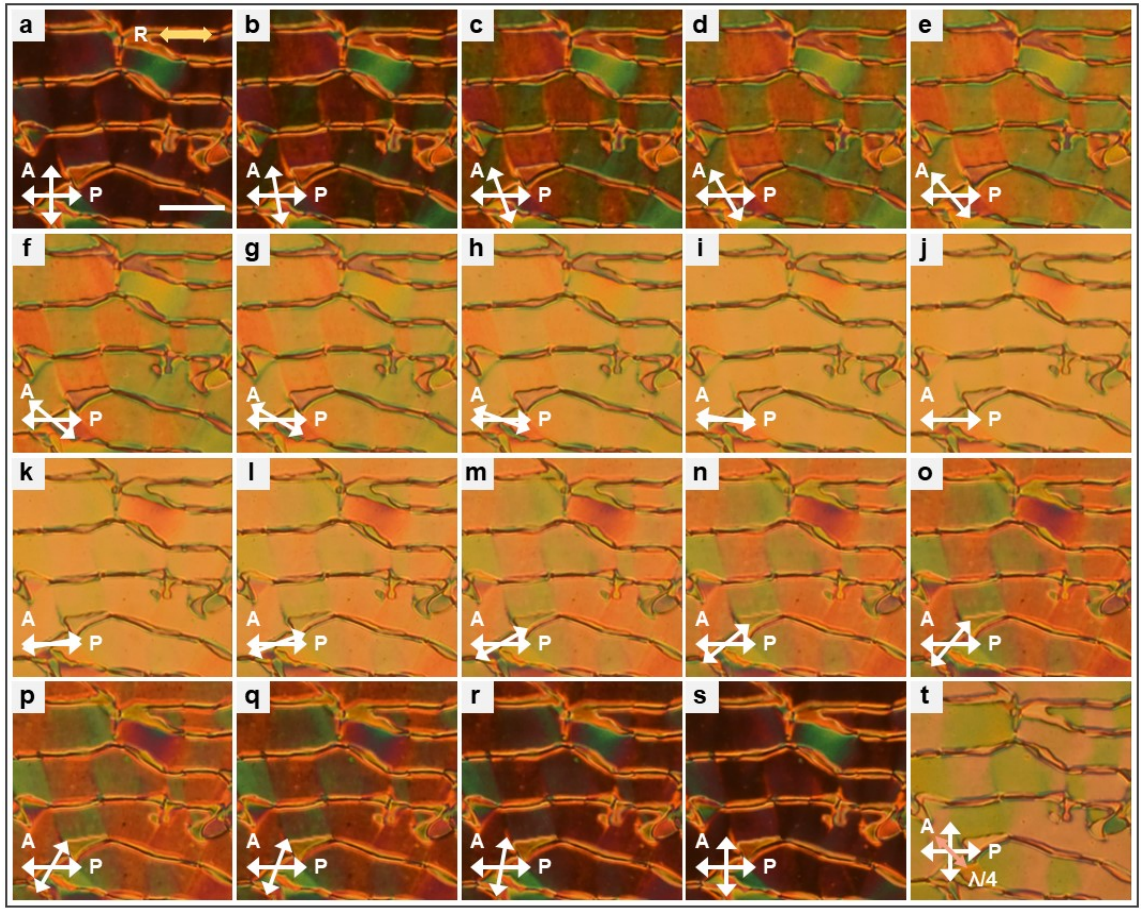

Figure **S14.** PLM observation of the $N_F$ band-like texture of RM-OC$_2$. **a**, The LC cell was placed parallel to the P polarizer under the crossed polarizers. **b-s**, The analyzer is rotated by 10°-180° counterclockwise. **t**, PLM images of $N_F$ band texture under crossed polarizers with a quarter-wave plate. The angle between the fast axis of the quarter-wave plate and the polarizer is 45°. The orange arrow indicates the fast axis of the quarter-wave plate. Optical graphs are taken in a parallel aligned cell. Rubbing direction is indicated by the yellow arrow. Scale bar: 50 μm. Temperature: 30°C.

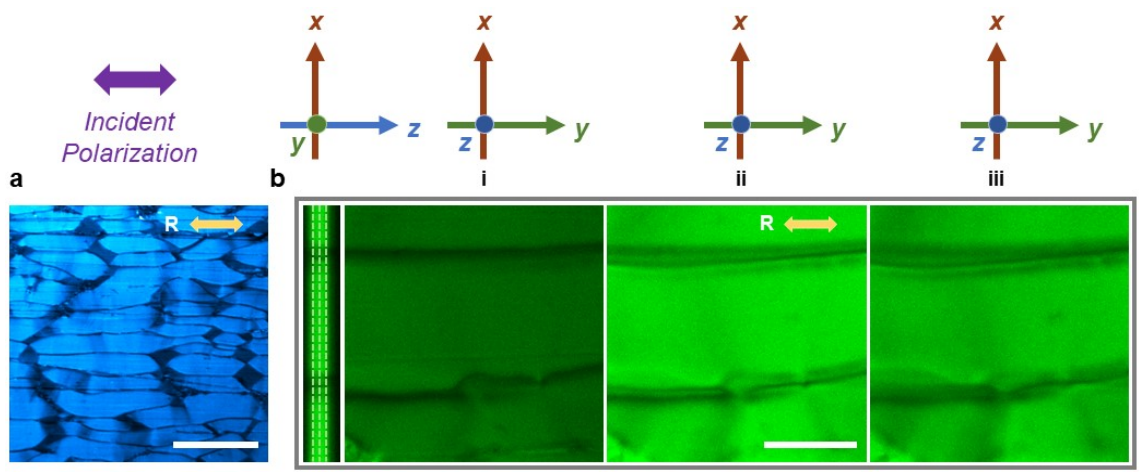

Figure **S15.** SHG confocal polarizing microscopy (SHG-CPM) image and FCPM images for $N_F$ band-like texture in a parallel aligned cell. The purple arrow represents the linear polarization of the incident light. Rubbing direction is indicated by the yellow arrow. **a**, xy-cross-sectional SHG-CPM images of the $N_F$ band-like texture of RM-OC$_2$. Scale bar, 50 μm. **b**, Cross-sectional FCPM images of the $N_F$ band-like texture of RM-OC$_2$. xy-cross-sectional images are visualized by a linearly polarization at three different of XZ-cross-sectional positions: positions i, ii, iii correspond to bottom surface, middle plane and top surface, respectively. Scale bars, 5 μm. Temperature: 30 °C.

*Supplementary Discussions 1-4*

*Supplementary Discussion 1.* The phase behaviors of RM-OC$_2$ are shown in Figures 2b-d and Figures S1c-j. The phase behavior upon the heating process is not simply mirror of that of the cooling process, exhibiting the phase sequence of N$_F$-[67.1 ˚C]-Iso-[82.1 ˚C]-unknown crystal X1-[105-120 ˚C]- unknown crystal X2-[140.1 ˚C]-Iso. During the temperature increases from 63 °C to 150 °C, the band-like texture of N$_F$ phase gradually melts into the Iso phase (Figs. S1c-f), and some crystal nuclei emerge. The crystalline phase X1 is replaced by another crystalline phase X2 at high temperatures, and then disappears upon further heating at 140.1 ˚C (Figs. S1g-j). The transition process suggests the metastability of the system. At low temperatures on cooling, though crystal is the most stable state, the slow nucleation and growth kinetics does not allow the system crystalized at the observation time scale (up to days). However, we can observe very small non-growing nucleus of crystal phase. Upon heating, the thermal perturbation provides fluctuation to the system to transit to the most-stable crystal phase. As shown in Figs. S1f-g, the domains readily crystallized upon heating can be attributed to the seeding effect from the remaining nucleus. This is also clearly observed in the DSC measurement (Fig. S1).

*Supplementary Discussion 2.* Figure 2a demonstrates the phase behaviors of RM-OC$_2$. The temperature of the Iso-N$_F$ transition is 64.6 °C and it exceeds the temperature limit of the oil immersion objective. Therefore, to obtain high resolution FCPM images, we choose an alternative material that shows the same N$_F$ droplets but at low temperatures. We synthesized two pure and stable conformers of DIO: *trans*- and *cis*- conformers, and mixed the *cis* conformer into the *trans* conformer. The synthesis of pure stereoisomers of DIO is described in Ref. 1. Under the mixing ratio of *trans*/*cis* conformers = 6/4, a direct Iso-N$_F$ phase transition occurs at 42 °C. The FCPM images of the N$_F$ droplets are captured at the temperature (Figs. S4b,c). The result is fully consistent with that of RM-OC2 (Fig. S3). This confirms the general incidence of the electric polarization meron-like structures.

*Supplementary Discussion 3.* As seen from the FCPM under a circular polarized pumping light, a decrease of fluorescence occurs near the droplet center, while the other areas show a nearly constant intensity of fluorescence. In the YZ-cross-sectional profile (Fig. S4c) the fluorescence intensity in the droplet center appears as a broaden black cylinder along the surface normal. This suggests the director field exhibits either a significant out-of-plane tilting towards the center (i.e. homeotropic-like in the center) or a disclination line. To determine the plausible structures, we compare the FCPM images of a disclination line with those of $N_F$ droplets. The line defect is constructed by employing a photo-alignment cell. The substrates are manufactured into patterns of +1 hedgehog-type defects. 5CB doped with Nile Red (c = 0.05 wt%) is filled into the cell. A +1 line disclination forms, connecting to both surfaces of the substrates. The dark region of the FCPM images of the line disclination (Fig. S4a) is apparently extremely sharp and thin even when one deliberately uses a low-NA objective. It is not consistent with the broaden black cylinder in $N_F$ droplets (Fig. S4b), suggesting that the disclination line is absent and the defects in the $N_F$ droplets. Therefore, at this stage, the concentric director fields might be either meron-like structure or escaped vortex structure.

To further differentiate the topology, we simulate the cross-sectional fluorescence profiles of these three structures and calculate the fluorescence intensity along paths in the center area and in the non-center areas (Figs. S5a-c). The same profiling treatments are employed to the experimental FCPM images to obtain the signal profile (Fig. S5d). From the signal profile of the escaped vortex (Fig. S5c), it is seen that, the signal profile along the center path is the thinnest, and outsides exhibit symmetrically identical signal distribution with respect to the center line. It is because the polarizations in the center line of the vortex lie in-plane without severe tilting. Predicted from the simulated meron-like structure (Figs. S9-10), as the singularity is missing in the droplet center, the decrease of the intensity in the droplet center is due to the out-of-plane tilting of the polarizations. Since the $N_F$-glass interface poses a degenerate anchoring, if the directors are closer to the substrates, the directors therefore exhibit less out-of-plane tilting so the fluorescence intensity decreases will be less in the area closer to the substrates than in the midplane. This feature is represented in the fluorescence cross section profile and signal profile of the meron-like structure: the extinction area in the droplet center is larger than that in the non-center area (Fig. S5b). Comparing the numerical FCPM images with the experimental results, it is clear that the $N_F$ droplets demonstrate the meron-like structure topology (Fig. S5d).

*Supplementary Discussion 4.* According to our observation, RM-OC$_2$ directly transits from isotropic to $N_F$ with an intermittent co-existence (Figs. 2b,c). When the sample is further cooled down, the droplets merge and transit to a band-like texture, where the line disclinations run mainly along the rubbing direction (Fig. 2d). The corresponding director and polarization field can be directly visualized by SHG microscopy. Fig. S15a demonstrates a large-area 2D SHG confocal polarizing microscopy (SHG-CPM) image of the band texture for the cell midplane during the cooling at 30 °C. In each domain, the SH signal shows the maximum intensity when the polarization is parallel to rubbing direction. This means the polarization points along the rubbing direction. Similar to Ref. [Li, J., Nishikawa, H., Kougo, J., *et al*. *Sci. Adv.* 2021, **7**(17): eabf5047.], the SHG interferometry reveals that the neighboring domains exhibit opposite polarity.